\documentclass[11pt,a4paper]{article}
\pdfoutput=1

\usepackage{alltt}
\usepackage{amsfonts}
\usepackage{asymptote}
\usepackage{bigstrut}
\usepackage{bm}
\usepackage{bbm}
\usepackage{booktabs}
\usepackage{calc}
\usepackage[makeroom]{cancel}
\usepackage[ntheorem]{empheq}
\usepackage[shortlabels,inline]{enumitem} 
\usepackage{esint}
\usepackage{fancyhdr}
\usepackage{footnote}
\usepackage{framed}
\usepackage{mathrsfs}
\usepackage{mathtools}
\usepackage{mflogo}
\usepackage[version=3]{mhchem}
\usepackage{multicol}
\usepackage{multirow}
\usepackage[framed,thmmarks,amsmath,hyperref]{ntheorem}
\usepackage{pdfrender}
\usepackage{rotating}
\usepackage{scalerel}
\usepackage{setspace}
\usepackage{siunitx}
\usepackage{slashed}
\usepackage{subfig}
\usepackage{tabularx}
\usepackage{tensor}
\usepackage{textcomp}
\usepackage{titlesec}
\usepackage{titling}
\usepackage[normalem]{ulem}
\usepackage{url}
\usepackage{verbatim}
\usepackage{wrapfig}
\usepackage{xparse}
\usepackage[vcentermath]{youngtab}

	\let\originalleft\left
	\let\originalright\right
	\renewcommand{\left}{\mathopen{}\mathclose\bgroup\originalleft}
	\renewcommand{\right}{\aftergroup\egroup\originalright}

		\DeclarePairedDelimiter\abs{\lvert}{\rvert}
		
		\renewcommand{\bar}[1]{\ensuremath\overline{#1}}

		\makeatletter
		\providecommand{\bigsqcap}{%
		  \mathop{%
		    \mathpalette\@updown\bigsqcup
		  }%
		}
		\newcommand*{\@updown}[2]{%
		  \rotatebox[origin=c]{180}{$\m@th#1#2$}%
		}
		\makeatother

		\newcommand*\diff{\mathop{}\!\mathrm{d}}

		\def\XXint#1#2#3{{\setbox0=\hbox{$#1{#2#3}{\int}$}
		    \vcenter{\hbox{$#2#3$}}\kern-.5\wd0}}

		\newcommand*{\R}{\mathbb{R}}
		\newcommand*{\Z}{\mathbb{Z}}
		
		\newcommand*{\F}{\mathbb{F}}
		\newcommand*{\N}{\mathbb{N}}

		\newcommand*{\bP}{\mathbb{P}}

		\newcommand*{\cN}{\mathcal{N}}
		\newcommand*{\cO}{\mathcal{O}}

		\protected\def\numpi{\text{\ensuremath{\pi}}}

		\DeclareMathOperator{\diag}{diag}

		\newcommand*{\Mod}[1]{\ensuremath{~(\mathrm{mod}~#1)}}

		\DeclareMathOperator{\SO}{SO}

		\DeclareMathOperator{\SU}{SU}
		\newcommand*{\su}{\ensuremath\mathfrak{su}}

		\DeclareMathOperator{\trace}{tr}

		\DeclareMathOperator{\U}{U}

		\newcommand*{\mat}[2][b]{\ensuremath\begin{#1matrix}#2\end{#1matrix}}
		
		\newcommand*{\needpic}[2][0.5]{\text{\textcolor{red}{\Large PUT A PICTURE HERE!!!}}\message{LaTeX Warning: Need picture!}}

		\def\DHLhksqrt#1#2{\setbox0=\hbox{$#1\sqrt{#2\,}$}\dimen0=\ht0
			\advance\dimen0-0.2\ht0
			\setbox2=\hbox{\vrule height\ht0 depth -\dimen0}%
			{\box0\lower0.4pt\box2}}
		
		\newcommand*{\ssyng}[1]{\text{\scriptsize$\yng(#1)$}}
		
		\newcommand*{\tyng}[1]{\text{\tiny$\yng(#1)$}}

	\theorembodyfont{\upshape}

	\theoremsymbol{\ensuremath{_\Box}}
	\newframedtheorem{Theorem}{Theorem}
	\newframedtheorem{Corollary}{Corollary}
	\newframedtheorem{Lemma}{Lemma}
	\newframedtheorem{Proposition}{Proposition}

	\theoremsymbol{}

	\theoremstyle{nonumberplain}
	\theoremheaderfont{\scshape}
	\theoremsymbol{\ensuremath{_\blacksquare}}
	
	\theoremstyle{nonumberbreak}
	
	\DeclareSIUnit\year{yr}
	\DeclareSIUnit\minute{min}
	\DeclareSIUnit\parsec{pc}
	\DeclareSIUnit\lightyear{lyr}

	\begin{asydef}
		import plain;
		import three;
		import feynman;
		usepackage("amsmath");
		usepackage("bm");
		usepackage("tensor");

		// set default line width to 0.8bp
			currentpen = linewidth(0.8);
		// scale all other defaults of the feynman module appropriately
			fmdefaults();
		// Turn off overpaint
			overpaint = false;

		// Definitions //
			//cap
				path LC = (0,0)..(-0.5,0.5)..(0,1);
				path RC = (0,0)..(0.5,0.5)..(0,1);

			//open end
				path LOEext = xscale(0.33)*LC;
				path LOEint = xscale(0.33)*RC;
				path ROEext = LOEint;
				path ROEint = LOEext;

			//legs
				//shifts and reflections
					transform LLxs = shift(-2.5, 0);
					transform RLxs = shift(2.5, 0);
					transform ULys = shift(0, 2);
					transform DLys = shift(0, -2);
					transform v = reflect((0,0),(1,0));
					transform h = reflect((0,0),(0, 1));

				path DLLoa = (-2.5,-2){right}..(-1.67,-1.625)..(-0.83,-0.875)..(0,-0.5){right};
				path DLLia = (-2.5,-1){right}..(-2,-0.75)..(-1.5,-0.25)..(-1,0);
				path[] DLLext = DLLoa^^DLLia^^DLys*LLxs*LOEext;
				path DLLint = DLys*LLxs*LOEint;

				path[] ULLext = v*DLLext;
				path ULLint = v*DLLint;

				path[] DRLext = h*DLLext;
				path DRLint = h*DLLint;

				path[] URLext = h*v*DLLext;
				path URLint = h*v*DLLint;

			//propagator
				path Pt = (0,0.5)--(1,0.5);
				path Pb = (0,-0.5)--(1,-0.5);
				path[] P = Pt^^Pb;

			//left pant
				path[] LPext = DLLext^^ULLext^^((-1,0))--(-0.85,0.04);
				path[] LPint = DLLint^^ULLint;

			//right pant
				path[] RPext = DRLext^^URLext^^((1,0))--(0.85,0.04);
				path[] RPint = DRLint^^URLint;

			//bubble
				//shift
					transform Bxs = shift(4, 0);

				path Bot = (0,0.5){right}..(2,1.25)..(4,0.5){right};
				path Bob = (0,-0.5){right}..(2,-1.25)..(4,-0.5){right};
				path Bit = (1.3,0)..(2,0.25)..(2.7,0);
				path Bib = (1.2,0.09275)..(1.3,0)..(2,-0.25)..(2.7,0)..(2.8,0.09275);
				path[] B = Bot^^Bob^^Bit^^Bib;
	\end{asydef}

\usepackage{jheppub}
\makeatletter
\g@addto@macro\bfseries{\boldmath}
\renewcommand{\@email}[1]{\texttt{#1}}
\def\NAT@sort{\z@}
\makeatother

\newcommand{\refsec}{Section}
\newcommand{\reftab}{Table}
\newcommand{\reffig}{Figure}
\newcommand{\refeqn}{Eq.}

\newcommand{\nd}{--}

\title{An infinite swampland of $\U(1)$ charge spectra in 6D supergravity
theories}
\author{Washington Taylor}
\author{and Andrew P. Turner}
\affiliation{
Massachusetts Institute of Technology, \\
77 Massachusetts Avenue, Cambridge, MA
}
\emailAdd{\tt{wati} \rm{at} \tt{mit.edu}}
\emailAdd{\tt{apturner} \rm{at} \tt{mit.edu}}
\preprint{MIT-CTP-4992}
\abstract{
We analyze the anomaly constraints on 6D supergravity theories with a single
abelian $\U(1)$ gauge factor. For theories with charges restricted to $q =
\pm1, \pm2$ and no tensor multiplets, anomaly-free models match those models
that can be realized from F-theory compactifications almost perfectly. For
theories with tensor multiplets or with larger charges, the F-theory
constraints are less well understood. We show, however, that there is an
infinite class of distinct massless charge spectra in the ``swampland'' of
theories that satisfy all known quantum consistency conditions but do not
admit a realization through F-theory or any other known approach to string
compactification. We also compare the spectra of charged matter in abelian
theories with those that can be realized from breaking nonabelian $\SU(2)$ and
higher rank gauge symmetries.
}
\keywords{abelian, anomaly cancellation, F-theory, swampland}

\begin{document}
\maketitle
\flushbottom

\section{Introduction}\label{sec:intro}

The largest spacetime dimension in which a supersymmetric theory can have
matter fields in any representation of the gauge group other than the adjoint
is six. This makes six dimensions a natural place to begin to try to
systematically understand the structure of matter in physical theories with
supersymmetry. Furthermore, in six dimensions, gravitational, gauge, and mixed
anomalies strongly constrain the gauge group and matter content
\cite{GreenSchwarzWest6DAnom,SagnottiGS}, and
additional quantum consistency constraints
\cite{SagnottiGS,KumarTaylorBound,SeibergTaylorLattices} further limit the set
of possible low-energy supergravity theories. The theories that satisfy these
constraints match fairly closely in certain regimes with the set of theories
that can be realized from string theory compactifications through F-theory,
suggesting that any consistent low-energy 6D supergravity theory may
arise from string theory \cite{KumarTaylorStringUni6D}. While there are still
a number of ways in which the set of theories that satisfy all known quantum
consistency theories is larger than the set of known string vacua,
understanding the ``swampland'' \cite{VafaSwamp,OoguriVafaSwamp} of apparently
consistent theories with no description in terms of any known class of string
theories has been a productive approach to developing our understanding of
both the set of quantum consistency conditions on gravity theories and the
structure of string theory vacua.

In this paper we investigate the constraints that 6D anomaly cancellation
conditions place on the charged matter content of theories with a single
$\U(1)$ gauge field. For theories with only nonabelian gauge fields, it is
known that (at least for theories with $T < 9$ tensor multiplets) there are a
finite number of distinct massless spectra of gauge fields and charged matter
representations that are consistent with anomaly cancellation and other simple
constraints such as the proper sign of the gauge kinetic term for all gauge
fields \cite{KumarTaylorBound,KumarMorrisonTaylorGlobalAspects}. One of the
goals of this paper is to inquire whether a similar finiteness bound can be
demonstrated for 6D theories with abelian gauge factors. The short answer is
that it seems that it cannot, at least without some further constraints. We
find that there indeed are infinite families of theories with distinct sets of
$\U(1)$-charged matter fields that appear consistent from anomaly cancellation
and other known constraints, even restricting to the simplest class of models
with $T = 0$ tensor multiplets. On the other hand, it is known that F-theory,
the most general approach known to constructing string vacua in six
dimensions, gives only a finite number of distinct possible massless spectra
\cite{KumarMorrisonTaylorGlobalAspects}. Thus, it seems that there is an
infinite swampland of $\U(1)$ theories in six dimensions.

While in this paper we focus primarily on quantum consistency constraints from
anomalies, independent of any UV completion of the theory, we also consider at
various places in the paper additional constraints that may limit the set of
theories from the F-theory point of view. In particular, for theories with an
abelian $\U(1)$ gauge group and matter restricted to charges $q = \pm1, \pm2$,
and for corresponding unHiggsed theories with $\SU(2)$ gauge groups, standard
F-theory constructions give constraints that are stronger than but similar in
structure to the constraints from anomaly cancellation. One particular
question that we address more generally for many of the anomaly-free $\U(1)$
spectra we find is whether a given spectrum of $\U(1)$-charged matter fields
can be realized by breaking through the Higgs mechanism a nonabelian theory
with $\SU(N)$ or other gauge group with matter charged in various
representations. This is a useful perspective in particular because nonabelian
theories are much better understood, both in terms of low-energy constraints
and in terms of F-theory realizations. In addition to providing insight on
swampland-type questions, understanding the allowed structure of abelian
matter charges and their relationship with nonabelian theories may also
provide new insight into the difficult problem of understanding how F-theory
solutions with $\U(1)$ factors and various matter charges can arise through
explicit Weierstrass model constructions.

Note that different authors use the term ``swampland'' in slightly different
ways. Here, to be precise, we define the swampland to be the set of theories
that obey all \emph{known} quantum consistency conditions yet cannot be
realized in any \emph{known} approach to string compactification. We focus
here on massless spectra of 6D $\cN = 1$ supergravity theories. Note that with
this definition, the swampland is a time-dependent class of theories;
discovery of a new quantum consistency condition or a new string construction
can reduce the swampland in scope. Note also that while for some spectra it
can be argued that no construction is possible in F-theory (as currently
formulated) or in any other known approach to 6D string compactification, for
other theories it is not known whether an F-theory construction is possible;
in the latter situation, we refer to such models here as ``possible''
swampland models, where the adjective ``possible'' reflects specifically on
our incomplete understanding of how to answer the well-defined mathematical
question of whether one of the finite set of distinct Weierstrass models over
an allowed F-theory base surface can realize a specific massless spectrum.

We systematically investigate the structure of 6D supergravity theories with a
single $\U(1)$ gauge field and their possible matter spectra. We begin with
some background on nonabelian and abelian anomaly conditions in 6D in
\refsec~\ref{sec:background}, where we characterize generic spectra for
theories with a single $\SU(N)$ or $\U(1)$ gauge group in terms of a pair of
parameters appearing in the anomaly cancellation conditions. In
\refsec~\ref{sec:relateAC}, we show that the $\SU(N)$ and $\U(1)$ anomaly
conditions can be directly related, giving closely parallel constraints to the
charged spectra. In \refsec~\ref{sec:charge2}, we focus on models with abelian
matter with charges $q = \pm1, \pm2$ only. We show that for theories with no
tensor multiplets ($T = 0$), the constraints from anomaly cancellation
directly match simple UV constraints from F-theory and there is no swampland
except for one specific model where the global structure of the gauge group
becomes relevant. For theories with more tensor multiplets the anomaly
cancellation conditions of the low-energy theory appear in a stronger form for
standard F-theory constructions, though this stronger form can be violated for
more exotic F-theory constructions. In \refsec~\ref{sec:asymptotics}, we
include charges $\abs{q} > 2$ and identify an infinite swampland of charge
spectra, even for theories without tensor multiplets. We then consider how the
number of anomaly-consistent spectra grows as the natural parameter from the
anomaly constraints increases. In \refsec~\ref{sec:largerCharge}, we consider
more explicitly theories with charge $q = \pm3, \pm4, \pm5$ matter, in order
to further understand the types of models that appear in the ``swampland.'' In
\refsec~\ref{sec:swampland}, we summarize our results on the spectra that
inhabit the swampland, and we finish with concluding remarks in
\refsec~\ref{sec:conclusion}.

As this work was being completed, we received a preliminary version of
\cite{MonnierMooreParkQuantization}, in which general constraints on 6D
supergravity theories from anomaly cancellation are analyzed, and which has
some overlap with the analysis of this paper --- in particular, some of the
conditions we identify and use in this paper are special cases of the more
general conditions analyzed in that paper.




\section{Background}\label{sec:background}

In six dimensions, gauge, gravitational, and mixed anomaly cancellation
imposes fairly strict constraints on the spectrum of gauge fields and matter
content that can be coupled to gravity in a consistent quantum theory
\cite{GreenSchwarzWest6DAnom,SagnottiGS}. We briefly review in this section
the local anomaly cancellation conditions for a 6D supergravity theory with a
single $\SU(N)$ or $\U(1)$ gauge field. We use these conditions to
parameterize the theories with the simplest, generic matter content, and
describe how theories containing matter with higher charges can be related to
generic models.

\subsection{Anomaly cancellation conditions}\label{subsec:background-AC}

For the purposes of this paper, we are interested in single $\SU(N)$ or
$\U(1)$ gauge group factors, so we will make use of the anomaly cancellation
(AC) conditions only for these cases.

\subsubsection{Anomalies for $\SU(N)$ models}
\label{subsubsec:background-AC-SUN}

Consider a single $\SU(N)$ gauge group factor, and let $x_R$ be the number of
hypermultiplets transforming in the irreducible representation $R$ of
dimension $d_R$. The numbers of massless vector multiplets and
hypermultiplets, denoted $V$ and $H$ respectively, are then given by
\begin{equation}
V = d_\text{Adj} = N^2 - 1\,, \quad H = \sum_R x_R d_R\,,
\end{equation}
The number of tensor multiplets is denoted by $T$.


The structure of the anomaly polynomial for 6D supergravity gives the
gravitational and nonabelian AC conditions
\cite{GreenSchwarzWest6DAnom,SagnottiGS}, which in the notation of
\cite{KumarMorrisonTaylorGlobalAspects} can be written as
\begin{subequations}
\label{eq:SUAC}
\begin{align}
273 &= H - V + 29 T\,, \label{eq:ACgrav} \\
a \cdot a &= 9 - T\,, \label{eq:SUACa} \\
a \cdot b &= -\frac{1}{6} \left(\sum_R x_R A_R - A_\text{Adj}\right)\,,
    \label{eq:SUACA} \\
0 &= \sum_R x_R B_R - B_\text{Adj}\,, \label{eq:SUACB} \\
b \cdot b &= \frac{1}{3} \left(\sum_R x_R C_R - C_\text{Adj}\right)\,.
    \label{eq:SUACC}
\end{align}
\end{subequations}
Here, $a$ and $b$ are vectors in a $(1 + T)$-dimensional real vector space
that carries a symmetric bilinear form $\Omega$ of signature $(1, T)$; the
notation $x \cdot y$ denotes the associated $\SO(1, T)$-invariant product
$\Omega_{\alpha \beta} x^\alpha y^\beta$. For each representation $R$ of
$\SU(N)$, the group theory coefficients $A_R$, $B_R$, and $C_R$ are defined by
\begin{equation}
\trace_R F^2 = A_R \trace F^2\,, \quad \trace_R F^4 = B_R \trace F^4
    + C_R \left(\trace F^2\right)^2\,,
\end{equation}
where $\trace$ denotes the trace in the fundamental representation and
$\trace_R$ denotes the trace in an arbitrary representation $R$. These group
theory coefficients can be computed for any $\SU(N)$ in a manner described in
\cite{Erler6DAnom,KumarTaylorBound,KumarParkTaylorT=0}, among other sources.
Values for several representations are given in \reftab~\ref{tab:ABC}.

Note that the anomaly equations really depend only on the local
structure of the gauge algebra. Thus, for example, the anomaly
conditions are the same for $\SU(2)$ and $\SO(3)$. For the most part
we ignore the global structure of the gauge group, but a case where
this distinction becomes relevant is discussed in
\refsec~\ref{subsec:charge2-T0}.

\begin{table}[!th]
\centering

\begin{tabular}{c>{$}c<{$}>{$}c<{$}>{$}c<{$}>{$}c<{$}} \toprule
$R$ & \text{Dimension} & A_R & B_R & C_R \\ \midrule
\tyng{1} & N & 1 & 1 & 0 \\[0.5em]
Adjoint & N^2 - 1 & 2 N & 2 N & 6 \\[0.5em]
\tyng{1,1} & \frac{N (N - 1)}{2} & N - 2 & N - 8 & 3 \\[0.5em]
\tyng{2,1} & \frac{N (N + 1) (N - 1)} {3} & N^2 - 3 & N^2 - 27 & 6 N \\[0.5em]
\tyng{2} & \frac{N (N + 1)}{2} & N + 2 & N + 8 & 3 \\[0.5em]
\tyng{3} & \frac{N (N + 1) (N + 2)}{6} & \frac{(N + 2) (N + 3)}{2}
    & \frac{N^2 + 17 N + 54}{2} & 3 (N + 4) \\[0.5em]
\tyng{4} & \frac{N (N + 1) (N + 2) (N + 3)}{24}
    & \frac{(N + 2) (N + 3) (N + 4)}{6}
    & \frac{(N + 4) \left(N^2 + 23 N + 96\right)}{6}
    & \frac{3 (N + 4) (N + 5)}{2} \\[0.5em] \bottomrule
\end{tabular}

\caption{Values of the group theory coefficients $A_R$, $B_R$, and $C_R$, and
the dimensions, associated with a variety of representations $R$ of $\SU(N)$,
for $N \ge 4$. For $\SU(2)$ and $\SU(3)$, no quartic Casimir exists, so $B_R =
0$. In this case, the value of $A_R$ is as given, and the value of $C_R$ is
obtained by taking $C_R + B_R / 2$ using the values given here.}
\label{tab:ABC}
\end{table}

\subsubsection{Anomalies for $\U(1)$ models}\label{subsubsec:background-AC-U1}

Consider now a single $\U(1)$ gauge factor. The irreducible representations
of $\U(1)$ have charges $q \in \Z$; we define $x_q$ to be the number of
hypermultiplets transforming in the charge-$q$ irreducible representation of
$\U(1)$. The anomaly polynomial in this case yields the abelian AC equations
\cite{Erler6DAnom,ParkTaylorAbelian},
\begin{subequations}
\label{eq:U1AC}
\begin{align}
a \cdot \tilde{b} &= -\frac{1}{6} \sum_{q > 0} x_q q^2\,,
    \label{eq:U1ACsqr} \\
\tilde{b} \cdot \tilde{b} &= \frac{1}{3} \sum_{q > 0} x_q q^4\,.
    \label{eq:U1ACquar}
\end{align}
\end{subequations}
Here, $\tilde{b}$ is once again an $\SO(1, T)$ vector. We sum over only $q >
0$ because each charge-$q$ hypermultiplet contains a field with charge $+q$
and a field with charge $-q$. Note that in these equations, $q = 1$ is not
necessarily the fundamental unit charge for the $\U(1)$. We discuss this
further in \refsec~\ref{subsec:background-addConstraints}.

\subsubsection{Geometric interpretation in F-theory}
\label{subsubsec:background-AC-geomInterp}

From the F-theory perspective, the quantities appearing in the nonabelian AC
conditions \eqref{eq:SUAC} have a geometric interpretation
\cite{SadovFTheoryGreenSchwarz,KumarMorrisonTaylorGlobalAspects}. For an
F-theory compactification on a base $B$ with canonical class $K$ and
nonabelian gauge group $\SU(N)$ associated with a codimension one fiber
singularity over a divisor in the homology class $\Sigma$ in $B$, we can
identify $a$ with the canonical class $K$ and $b$ with the divisor class
$\Sigma$. In this case, the $\SO(1, T)$ product is identified with the
intersection product on divisor classes. Thus, we have
\begin{subequations}
\begin{align}
a \cdot a &= K \cdot K\,, \\
a \cdot b &= K \cdot \Sigma\,, \\
b \cdot b &= \Sigma \cdot \Sigma\,.
\end{align}
\end{subequations}
Defining the self-intersection of $\Sigma$ to be $n := \Sigma \cdot \Sigma$,
we can then use the Riemann\nd{}Roch formula,
\begin{equation}
2 (g - 1) = \Sigma \cdot (K + \Sigma) = n + K \cdot \Sigma\,,
\end{equation}
where $g$ is the genus of curve $\Sigma$, to rewrite \eqref{eq:SUAC} as
\begin{subequations}
\label{eq:geomSUAC}
\begin{align}
273 &= H - V + 29 T\,, \label{eq:geomSUACgrav} \\
K \cdot K &= 9 - T\,, \label{eq:geomSUACK} \\
2 (g - 1) - n &= -\frac{1}{6} \left(\sum_R x_R A_R - A_\text{Adj}\right)\,,
    \label{eq:geomSUACA} \\
0 &= \sum_R x_R B_R - B_\text{Adj}\,, \label{eq:geomSUACB} \\
n &= \frac{1}{3} \left(\sum_R x_R C_R - C_\text{Adj}\right)\,.
    \label{eq:geomSUACC}
\end{align}
\end{subequations}

When the gauge group is abelian and the only gauge group factor is a single
$\U(1)$, the geometric interpretation of $\tilde{b}$, as elaborated in
\cite{MorrisonParkU1}, is
\begin{equation}
\label{eq:MP-b}
\tilde{b} = 2 (-K + [z])\,,
\end{equation}
where $[z]$ is the divisor class of the $z$-component of the section that
generates the Mordell\nd{}Weil group of rational sections of the elliptic
fibration. In an F-theory model, therefore, $\tilde{b}$ must be both effective
and even. In analogy with the nonabelian case, we define $\tilde{n}, \gamma$
through
\begin{equation}
\tilde{b} \cdot \tilde{b} = \tilde{n}\,, \quad a \cdot \tilde{b} = -\gamma\,.
\end{equation}
The abelian AC conditions \eqref{eq:U1AC} then become
\begin{subequations}
\label{eq:geomU1AC}
\begin{align}
6 \gamma &= \sum_{q > 0} x_q q^2\,, \\
3 \tilde{n} &= \sum_{q > 0} x_q q^4\,.
\end{align}
\end{subequations}
A unified geometric description of the abelian and nonabelian anomaly
conditions in the framework of F-theory is given in \cite{ParkAnomalies}.

\subsection{Additional constraints}\label{subsec:background-addConstraints}

In addition to the local anomaly cancellation conditions, we trivially have
integrality and non-negativity conditions on $x_R, x_q$: the number of
hypermultiplets transforming in a given representation must be a non-negative
integer. The exception is in the case of quaternionic representations of a
nonabelian gauge group. We can have so-called ``half-hypermultiplets''
transforming in these representations, which have half the field content of a
full hypermultiplet; a full hypermultiplet transforming in the representation
$R$ contains an equal number of fields transforming in both $R$ and in
$\bar{R}$, while a half-hypermultiplet has half the field content, all
transforming only under $R$. For our purposes, this amounts to allowing $x_R$
to be a (non-negative) half-integer for such representations. (Note that all
even-dimensional irreducible representations of $\SU(2)$ are quaternionic.)

In addition to the local anomaly cancellation conditions we have described
above, there are also global anomalies that must cancel. In particular
\cite{WittenSU2Anom,BershadskyVafaAnom6D,SuzukiTachikawaAnom6D,
KumarMorrisonTaylor6DSUGRA,MonnierMooreParkQuantization}, for $\SU(2)$ and
$\SU(3)$ theories with only fundamental and adjoint matter fields, the
anomalies will cancel if the following conditions are satisfied:
\begin{equation}
\label{eq:globalAnom}
\begin{alignedat}{2}
\SU(2)\colon&& \quad x_{\,\tyng{1}} + 4 x_{\, \tyng{2}} &= 4 \Mod{6}\,, \\
\SU(3)\colon&& \quad x_{\,\tyng{1}} &= 0 \Mod{6}\,.
\end{alignedat}
\end{equation}
These conditions are automatically satisfied by the generic spectra described
in the next section. There are no global anomalies for larger $\SU(N)$ groups.
Since, as described in \cite{WittenSU2Anom}, the global anomaly constraints
depend only on the net contribution from all matter representations to the
group theory coefficients $A_R, B_R, C_R$, we expect that the global anomaly
constraints are therefore also satisfied in all theories we consider here that
satisfy the local anomaly cancellation conditions. Cancellation of the local
anomalies guarantees the cancellation of global anomalies if $b$ is an element
of the string charge lattice $\Gamma$ --- a condition discussed below.

We now consider integrality constraints on the parameters $n, g, \tilde{n},
\gamma$, and on $a, b, \tilde{b}$. We begin with the nonabelian parameters. By
simply using the properties of the group theory coefficients $A_R, B_R, C_R$,
it was shown in \cite{KumarMorrisonTaylorGlobalAspects} that $n = b \cdot b$
and $a \cdot b$ must take integer values in any theory satisfying the AC
conditions. Thus, $n$ must be an integer for any anomaly-free theory. A
similar argument \cite{MonnierMooreParkQuantization} shows that $a \cdot b + b
\cdot b \in 2 \Z$ directly from the group theory coefficients, so that $g$ is
always an integer. In \cite{SeibergTaylorLattices}, it was shown that the
lattice $\Gamma$ of string charges coupled to the two-form fields of the
theory, with Dirac product defined by $\Omega$, is a unimodular lattice. The
structure of the Green\nd{}Schwarz terms in the 6D supergravity action
suggests that $a, b$, which are a priori simply vectors in the real vector
space $\R^{1, T}$ with inner product $\Omega$, must in fact be elements of the
string charge lattice $\Gamma$, as they are associated with the string charges
of gauge and gravitational instantons, respectively; a similar consideration
in the abelian sector suggests that $\tilde{b}$ should also be an element of
the lattice $\Gamma$, and hence that $\tilde{n}$ and $\gamma$ are integers,
although the anomaly conditions only directly constrain, e.g., that $3
\tilde{n}$ is an integer. (For example, the abelian equations are solved by
$\tilde{n} = 19 / 3, \gamma = 7 / 6, x_1 = 3, x_2 = 1$, although there is no
theory with an integral lattice $\Gamma$ and $\tilde{b} \in \Gamma$ that gives
these values). The string charge lattice also has associated with it a
positive cone supporting BPS strings, and we expect that $-a$, $b$, and
$\tilde{b}$ lie in this positive cone. The consistency constraints on this
positive cone are not well understood from the supergravity point of view,
although in this paper this condition is primarily only relevant in the
discussion of $T = 1$ models in \refsec~\ref{subsec:charge2-T1}. Note that in
most of the examples we consider here, we take $T = 0$, where, as discussed in
\refsec~\ref{subsec:charge2-T0}, $\tilde{b}$ automatically takes (even)
integer values for models with generic abelian charges $\pm1, \pm2$, and the
positivity condition that $-a, b, \tilde{b} > 0$ also arises naturally.

Finally, note that for every solution $a, \tilde{b}, x_q$ of the abelian AC
equations~\eqref{eq:U1AC}, there is an infinite family of solutions $a' = a,
\tilde{b}' = k^2 \tilde{b}, x_{k q}' = x_q$, for $k = 1, 2, \dotsc$. At the
level of massless spectra, these solutions may seem equivalent up to a scaling
of the overall charge, but because $\tilde{b}$ has an absolute scaling
relative to the lattice $\Gamma$, these models must be considered as
physically distinct. A distinct question is the fundamental unit of charge
under the $\U(1)$ gauge group. Even though the models with e.g. $\tilde{b}' =
4 \tilde{b}, x_{2 q}' = x_q$ are physically distinct, the fundamental unit of
charge in the $\tilde{b}', x'$ theory may naturally be taken to be $q = 2$, or
may be $q = 1$, in particular in a scenario where there are massive states in
the spectrum with that charge. We give an explicit example of a pair of
distinct physical theories with charges related by scaling in this way in
\refsec~\ref{subsec:charge2-T0}, and return to this general question in the
discussion of the infinite apparent swampland for $\U(1)$ models later in the
paper.

\subsection{Generic and non-generic $\SU(N)$ models}
\label{subsec:background-genericSUN}

We can now determine all anomaly-free models for an $\SU(N)$ gauge theory that
contains only fundamentals, adjoints, and two-index antisymmetric matter
fields (in the case of $\SU(2)$ and $\SU(3)$, which have no quartic Casimir,
we consider only fundamentals and adjoints). In terms of the self-intersection
$n$ and genus $g$ as defined above, solving the $\SU(N)$ AC conditions
\eqref{eq:geomSUAC} yields models of the form
\begin{subequations}
\label{eq:geomSUACsol}
\begin{align}
[6 n + 2 (N + 6) (1 - g)] \times \ssyng{1} + g \times \text{Adj}\,,
    \quad N &= 2, 3\,, \\
[(8 - N) n + 16 (1 - g)] \times \ssyng{1} + g \times \text{Adj}
    + \left[n + 2(1 - g)\right] \times \ssyng{1,1}\,, \quad N &\ge 4\,.
\end{align}
\end{subequations}
These are in several senses the most generic types of $\SU(N)$ spectra for 6D
supergravity theories. For any values of $n, g$ for which this massless
spectrum is possible (non-negative), the resulting model will have more
uncharged scalar fields than any other mass spectrum satisfying anomaly
cancellation with the same values of $n, g$. This can be understood in
F-theory from the fact that the representations used here come from the most
generic types of codimension two singularities; in general, more exotic
representations involve tuning to special points in the moduli space where the
number of uncharged scalar fields is smaller
\cite{MorrisonTaylorMaS,AndersonGrayRaghuramTaylorMiT,KleversMorrisonRaghuramTaylorExotic}.
Note, however, that for some values of $n, g$ some of the representations in
these models have negative multiplicity; in such cases, this ``generic'' type
of matter is not possible and if there are solutions to the anomaly equations
with non-negative multiplicities $x_R$ they must include other matter
representations.

From the generic models \eqref{eq:geomSUACsol}, we can determine the other
possible spectra for given values of $n, g$. The group coefficients $A_R$,
$B_R$, and $C_R$ for larger representations will be given by a linear
combination of those for the fundamental, adjoint, and antisymmetric, so we
can always exchange this linear combination of representations in
\eqref{eq:geomSUACsol} for the corresponding representation $R$, and the
resulting spectrum will automatically satisfy the AC conditions.

For example, consider the $\ssyng{3}$ representation of $\SU(2)$; the anomaly
coefficients of this representation can be related to those of the generic
matter representations by a linear system of equations,
\begin{subequations}
\begin{align}
A_{\,\tyng{3}} &= c_1 A_{\,\tyng{1}} + c_2 A_{\,\tyng{2}}\,, \\
C_{\,\tyng{3}} &= c_1 C_{\,\tyng{1}} + c_2 C_{\,\tyng{2}}\,.
\end{align}
\end{subequations}
Using the values from \reftab~\ref{tab:ABC} gives $c_1 = -14$ and $c_2 = 6$;
thus, we can freely make the exchange
\begin{equation}
\label{eq:SU2sym3exchange}
14 \times \bm{1} + 6 \times \ssyng{2} \longleftrightarrow
    1 \times \ssyng{3} + 14 \times \ssyng{1}
\end{equation}
in any $\SU(2)$ model satisfying the AC conditions to yield another model
satisfying the AC conditions. Note that we must check that the non-negativity
constraint is satisfied after the exchange, but note also that in some cases
such an exchange may take an inconsistent model, such as one with more than 6
fields in the $\ssyng{2}$ representation but a negative number of fields in
the $\ssyng{1}$ representation, to a consistent model.

Representations that are related in this way are termed \emph{anomaly
equivalent} \cite{MorrisonTaylorMaS,GrassiMorrisonAnomalies}. Here we refer to
the transformation between one model and another using anomaly equivalent
representations as an \emph{exchange}. Note that these exchanges need not be
realizable by any physical process; here we simply use these exchanges as an
organizational tool for classifying solutions to the anomaly equations. There
are many instances, however, in which these exchanges can actually be realized
as physical transitions
\cite{AndersonGrayRaghuramTaylorMiT,KleversMorrisonRaghuramTaylorExotic}.

\subsection{Generic and non-generic $\U(1)$ models}
\label{subsec:background-genericU1}

We can similarly compute all anomaly-free models for a $\U(1)$ gauge theory
that contains only charges $q = \pm1, \pm2$. Solving the $\U(1)$ AC conditions
\eqref{eq:geomU1AC} yields models of the form
\begin{equation}
\label{eq:geomU1ACsol}
\left(8 \gamma - \tilde{n}\right) \times (\bm{\pm1})
    + \left(\frac{\tilde{n} - 2 \gamma}{4}\right) \times (\bm{\pm2})\,,
\end{equation}
where $(\bm{\pm n})$ indicates a representation with charge $\pm n$.

As in the $\SU(N)$ case, we can, in principle, use this result to construct
all other anomaly-free $\U(1)$ models by way of exchanges: comparing
\eqref{eq:geomU1AC} and \eqref{eq:geomSUAC}, we see that the $\U(1)$ analogues
of $A_R$ and $C_R$ are simply $q^2$ and $q^4$, allowing us to determine an
anomaly equivalence between any higher charge and a linear combination of
fields of charges $\pm1, \pm2$. For charge $q = \pm3$, for example, solving
the system
\begin{subequations}
\begin{align}
3^2 &= c_1 1^2 + c_2 2^2\,, \\
3^4 &= c_1 1^4 + c_2 2^4\,,
\end{align}
\end{subequations}
gives $c_1 = -15, c_2 = 6$, so we can freely make the exchange
\begin{equation}
\label{eq:U1charge3exchange}
10 \times (\bm{0}) + 6 \times (\bm{\pm2}) \longleftrightarrow
    1 \times (\bm{\pm3}) + 15 \times (\bm{\pm1})
\end{equation}
in an anomaly-free $\U(1)$ model to yield another anomaly-free $\U(1)$ model
(again, subject to non-negativity). As with the nonabelian theories, we treat
these exchanges as a formal way of relating low-energy theories with distinct
spectra, though we expect that in many cases there are corresponding physical
transitions (see, e.g., the discussion in \cite{Raghuram34}).

Note that when $T \ge 9$, there are possible values of $a, \tilde{b}$ that
satisfy $a \cdot \tilde{b} = \tilde{b} \cdot \tilde{b} = 0$. These correspond
to non-Higgsable $\U(1)$ models with no matter content, and $\tilde{n} =
\gamma = 0$. Explicit examples of such theories have been identified in
F-theory \cite{MartiniTaylorSemitoric,MorrisonParkTaylorNHAbelian}. Such
models are generally associated with Higgsing $\SU(2)$ models that have only a
single adjoint representation; for example, this describes any F-theory model
with $T = 9$ where the gauge factor is supported on a curve in the genus one
class $\Sigma = -K$.

\section{Relating the $\U(1)$ and $\SU(N)$ anomaly cancellation equations}
\label{sec:relateAC}

Since nonabelian gauge groups and matter are easier to understand and
constrain than abelian gauge groups and matter, both from the low-energy point
of view and from F-theory, it is often helpful to relate allowed abelian
structures to nonabelian models that can be broken to reproduce abelian gauge
groups and matter through Higgsing (see, e.g.,
\cite{MorrisonParkU1,MorrisonTaylorSections,CveticKleversPiraguaTaylorU1U1}).
In particular, if a nonabelian gauge group and matter spectrum can be realized
in F-theory and leads in field theory to a given abelian gauge group and
matter spectrum after a Higgsing process, then there must also be an F-theory
realization of the resulting abelian theory (though the explicit realization
of the Higgsing process in an F-theory Weierstrass model can be rather tricky
\cite{Raghuram34}). Since there are strict bounds on the finite set of
nonabelian gauge groups and matter that can be realized in F-theory, only a
limited set of abelian models can be unHiggsed to a nonabelian model with an
F-theory realization. Note that there are also $\U(1)$ models that can be
realized in F-theory that cannot be unHiggsed to a good $\SU(2)$ model in
F-theory; we encounter some examples of this later in the paper. Note also
that there are only a finite number of massless spectra with a single $\SU(N)$
gauge factor that satisfy the anomaly conditions for any value of $T$, since
the number of hypermultiplets is bounded above for any given $N$ in such a
situation, so only a finite number of $\U(1)$ models can be unHiggsed to an
anomaly-free $\SU(N)$ theory.

In this section, we directly relate the anomaly cancellation conditions and
the structure of matter spectra in theories with a $\U(1)$ gauge group to the
AC conditions and matter spectra in theories with an $\SU(N)$ gauge group, and
we describe a general class of Higgs transitions that break $\SU(N)$ to
$\U(1)$.

We first compare the AC conditions for the $\SU(N)$ and $\U(1)$ cases
directly. The $\U(1)$ AC conditions \eqref{eq:U1AC} for charges $q = \pm1,
\pm2$ are
\begin{subequations}
\label{eq:genericU1AC}
\begin{align}
-6 a \cdot \tilde{b} &= x_1 + 4 x_2\,, \\
3 \tilde{b} \cdot \tilde{b} &= x_1 + 16 x_2\,.
\end{align}
\end{subequations}
If we do not impose non-negativity conditions, then any anomaly-free $\U(1)$
model can be exchanged to a model with only these charges, and the matter
content will satisfy these two constraints.

Similarly, any $\SU(N)$ model can be exchanged to a model with only
fundamentals, adjoints, and antisymmetrics if we do not impose the
non-negativity conditions. For such a model, the $\SU(N)$ AC conditions
\eqref{eq:SUAC} yield
\begin{subequations}
\label{eq:genericSUAC}
\begin{align}
-6 a \cdot b &= x_{\,\tyng{1}} + 2 N (x_\text{Adj} - 1)
    + (N - 2) x_{\,\tyng{1,1}}\,, \label{eq:genericSUACa} \\
3 b \cdot b &= \frac{1}{2} x_{\,\tyng{1}} + (N + 6) (x_\text{Adj} - 1)
    + \frac{1}{2} (N - 2) x_{\,\tyng{1,1}}\,, \label{eq:genericSUACb}
\end{align}
\end{subequations}
for $N > 3$. In the case of $N = 2, 3$, we take $x_{\,\tyng{1,1}} = 0$ here;
for $N > 3$, \eqref{eq:genericSUACb} follows from adding
\refeqn~\eqref{eq:SUACC} to half of \refeqn~\eqref{eq:SUACB}.

We now assume that $\tilde{b} = k b$, for $k \in \Z$. In such circumstances,
we can equate the right-hand sides of \eqref{eq:genericU1AC} with those of
\eqref{eq:genericSUAC}, appropriately scaled, yielding
\begin{subequations}
\begin{align}
x_1 + 4 x_2 &= k x_{\,\tyng{1}} + 2 k N \left(x_\text{Adj} - 1\right)
    + k (N - 2) x_{\,\tyng{1,1}}\,, \\
x_1 + 16 x_2 &= \frac{k^2}{2} x_{\,\tyng{1}}
    + k^2 (N + 6) \left(x_\text{Adj} - 1\right)
    + \frac{k^2}{2} (N - 2) x_{\,\tyng{1,1}}\,.
\end{align}
\end{subequations}
From these equations, we find
\begin{subequations}
\label{eq:1-n-relationship}
\begin{align}
x_1 &= \frac{k}{6} (8 - k) x_{\,\tyng{1}}
    + \frac{k}{3} [8 N - k (N + 6)] \left(x_\text{Adj} - 1\right)
    + \frac{k}{6} (8 - k) (N - 2) x_{\,\tyng{1,1}} \,, \\
x_2 &= \frac{k}{24} (k - 2) x_{\,\tyng{1}}
    + \frac{k}{12} [k (N + 6) - 2 N] \left(x_\text{Adj} - 1\right)
    + \frac{k}{24} (k - 2) (N - 2) x_{\,\tyng{1,1}}\,.
\end{align}
\end{subequations}

By inspection, if $k$ is $0$ or $2$ modulo $6$, then $x_1, x_2 \in \Z$ for any
$N, x_{\,\tyng{1}}, x_\text{Adj}, x_{\,\tyng{1,1}} \in \Z$. We thus expect $k
\equiv 0, 2 \Mod 6$, for any valid Higgsing from $\SU(N)$ to $\U(1)$ that
gives $\tilde{b} = k b$ for some $k$. Note that the relationship
\eqref{eq:1-n-relationship} was found by considering only the AC conditions,
without any reference to the explicit symmetry-breaking pattern.

For an $\SU(N)$ gauge theory, we can explicitly construct a Higgsing pattern
to $\U(1)$ with $\tilde{b} = m (m - 1) b$, for $2 \le m \le N$. To see this,
first note that one possible Higgsing of an $\SU(N)$ to a $\U(1)$ using two
adjoints is achieved by giving a VEV to a Cartan generator in one adjoint
matter field that breaks $\SU(N) \to \U(1)^{N - 1}$, and then giving VEVs to
$N - 2$ of the $\U(1)$ charges coming from the second adjoint of $\SU(N)$ in
order to break $N - 2$ of the remaining $\U(1)$ factors. Without loss of
generality, let the surviving $\U(1)$ generator be $\diag(1, 1, \dotsb, 1, -N
+ 1)$; then we see that under this Higgsing, the fundamental, adjoint, and
antisymmetric representations decompose as
\begin{equation}
\label{eq:genericSUdecomp}
\begin{aligned}
\ssyng{1} &\to (N - 1) \times (\bm{\pm 1}) + 1 \times (\bm{\pm(N - 1)})\,, \\
\text{Adj} &\to (N - 1)^2 \times (\bm{0})
    + 2 (N - 1) \times (\bm{\pm N})\,, \\
\ssyng{1,1} &\to \frac{(N - 1) (N - 2)}{2} \times (\bm{\pm2})
    + (N - 1) \times (\bm{\pm(N - 2)})\,.
\end{aligned}
\end{equation}
We can then determine the relationship between $\tilde{b}$ and $b$ by
comparing the AC equations for the two theories. After the Higgsing, the
resulting $\U(1)$ model must satisfy
\begin{equation}
\label{eq:decompU1ACa}
-6 a \cdot \tilde{b} = x_1 + 4 x_2 + (N - 2)^2 x_{N - 2} + (N - 1)^2 x_{N - 1}
+ N^2 x_N\,,
\end{equation}
and we see from \refeqn~\eqref{eq:genericSUdecomp} that
\begin{equation}
\begin{gathered}
x_1 = (N - 1) x_{\,\tyng{1}}\,, \quad
    x_2 = \frac{(N - 1) (N - 2)}{2} x_{\,\tyng{1,1}}\,, \\
x_{N - 2} = (N - 1) x_{\,\tyng{1,1}}\,, \quad
    x_{N - 1} = x_{\,\tyng{1}}\,, \quad
    x_N = 2 (N - 1) (x_\text{Adj} - 1)\,,
\end{gathered}
\end{equation}
noting that Higgsing on an adjoint charge uses up one of the adjoints.
Plugging these into \refeqn~\eqref{eq:decompU1ACa} and comparing with
\refeqn~\eqref{eq:genericSUACa} yields
\begin{equation}
-6 a \cdot \tilde{b} = N (N - 1) \left[x_{\,\tyng{1}} + 2 N (x_\text{Adj} - 1)
    + (N - 2) x_{\,\tyng{1,1}}\right] = N (N - 1) (-6 a \cdot b)\,.
\end{equation}
Thus, this Higgsing gives $\tilde{b} = N (N - 1) b$.

We could instead achieve $\tilde{b} = (N - 1) (N - 2) b$ by first Higgsing on
an adjoint charge that breaks $\SU(N) \to \SU(N - 1) \times \U(1)$, and then
carrying out the Higgsing described above. By iterating this step, we can thus
achieve Higgsings of $\SU(N) \to \U(1)$ with $\tilde{b} = m (m - 1) b$ for any
$2 \le m \le N$. Note that, as anticipated above, $m (m -1) \equiv 0, 2 \Mod
6$ for any $m$.

This analysis gives us a way of relating a broad class of generic $\U(1)$
theories to Higgsed $\SU(N)$ models with generic matter spectra.

\section{Charge $\abs{q} \le 2$}\label{sec:charge2}

We now consider the explicit classification of $\U(1)$ models and their
unHiggsings. We begin by considering models with $\U(1)$ charges $\abs{q} \le
2$, i.e., $x_q = 0$ for $\abs{q} \ge 3$. For arbitrary $T$, models will be
parameterized by $\gamma, \tilde{n}$ in the case of $\U(1)$ models and $g, n$
in the case of $\SU(2)$ models, where these parameters are constrained by the
condition that the numbers of matter fields $x_q, x_R$ in any consistent model
are non-negative.

\subsection{$\U(1)$ models}\label{subsec:charge2-U1}

Restricting ourselves to $\abs{q} \le 2$, we have $\U(1)$ models of the form
\eqref{eq:geomU1ACsol},
\begin{equation}
\left(8 \gamma - \tilde{n}\right) \times (\bm{\pm1})
    + \left(\frac{\tilde{n} - 2 \gamma}{4}\right) \times (\bm{\pm2})\,.
\end{equation}
In this case, we have $H = x_0 + x_1 + x_2$, with $x_0$ the number of trivial
representations, and $V = 1$, so the gravitational condition \eqref{eq:ACgrav}
becomes
\begin{equation}
274 = x_0 + x_1 + x_2 + 29 T\,.
\end{equation}
Considering only the charged matter, this yields the condition
\begin{equation}
274 \ge 29 T - \frac{3 (\tilde{n} - 10 \gamma)}{4}\,.
\end{equation}

The anomaly equations and integrality/non-negativity of charges impose the
conditions
\begin{equation}
\label{eq:q2-conditions}
3 \tilde{n} \in \Z\,, \quad \tilde{n} - 2 \gamma \in 4 \Z\,,
    \quad 2 \gamma \le \tilde{n} \le 8 \gamma\,,
\end{equation}
though, as discussed in section \ref{subsec:background-addConstraints},
$\tilde{n}$ is presumably constrained to be an integer in any
quantum-consistent low-energy theory. Note that the second of these conditions
matches the constraint numbered 5 in \cite{MonnierMooreParkQuantization} in
the case of a single abelian gauge factor. Note also that the last condition
in \eqref{eq:q2-conditions} can be written as
\begin{equation}
\label{eq:q2-last}
-2 a \cdot \tilde{b} \le \tilde{b} \cdot \tilde{b} \le -8 a \cdot \tilde{b}\,.
\end{equation}
This form of the constraint will be helpful in connecting to a related
stronger condition on certain classes of F-theory models in
\refsec~\ref{subsec:charge2-fTheory}. Note that we can also derive the
constraints \eqref{eq:q2-last} directly from \eqref{eq:genericU1AC}; the first
bound follows from the condition $4 x_2 \le 16 x_2$, and the second can be
found by multiplying the first equation by 4 and noting that $4 x_1 \ge x_1$.

\subsection{$\SU(2)$ models}\label{subsec:charge2-SU2}

As we have seen, the anomaly-free $\SU(N)$ models with only generic matter
representations are given by \eqref{eq:geomSUACsol}. Specifically, for
$\SU(2)$ the anomaly-free models are of the form
\begin{equation}
\label{eq:SU2charge2}
[6 n + 16 (1 - g)] \times \ssyng{1} + g \times \ssyng{2}\,.
\end{equation}
We can obtain models containing larger representations by using the group
theory coefficients to exchange these additional representations for some
number of fundamentals and adjoints. However, models with only fundamentals
and adjoints will Higgs down to $\U(1)$ models with charges $q = \pm1, \pm2$,
which is the current case of interest. We have $H = x_0 + 2 x_{\,\tyng{1}} + 3
x_{\,\tyng{2}}$ 
and $V = 3$, so the gravitational condition \eqref{eq:ACgrav} becomes
\begin{equation}
276 \ge 2 x_{\,\tyng{1}} + 3 x_{\,\tyng{2}} + 29 T\,,
\end{equation}
or
\begin{equation}
244 \ge 12 n + 29 (T - g)\,.
\end{equation}

We can now Higgs these down to $\U(1)$ models. We Higgs on an adjoint charge
that leaves the generator $T_3$ unbroken (by this, we mean that we give an
adjoint hypermultiplet a VEV equal to this adjoint charge). The resulting
$\U(1)$ charges are twice the eigenvalues of the generator $T_3$ in the given
representation. Thus,
\begin{subequations}
\label{eq:breaking-symmetric}
\begin{align}
\ssyng{1} &\to 2 \times (\bm{\pm1})\,, \\
\ssyng{2} &\to 1 \times (\bm{0}) + 2 \times (\bm{\pm2})\,.
\end{align}
\end{subequations}
Higgsing uses up one of the adjoints, so we find models of the form
\begin{equation}
[12 n + 32 (1 - g)] \times (\bm{\pm1}) + 2 (g - 1) \times (\bm{\pm2})\,.
\end{equation}
Matching these with the models from \refsec~\ref{subsec:charge2-U1}, we find
\begin{equation}
\label{eq:SU2-matching}
\tilde{n} = 4 n\,, \quad \gamma = 2 n + 4 (1 - g)\,.
\end{equation}
Note that $\tilde{n} = 4 n$ corresponds to $\tilde{b} = 2 b$, which agrees
with our expectation from \refsec~\ref{sec:relateAC} for an $\SU(2)$ Higgsing.

We know that both $n$ and $g$ are integers such that the number of each
representation in \eqref{eq:SU2charge2} is non-negative (the number of each
representation is always integral due to the integrality of $n$ and $g$). By
contrast, for the abelian theories we only know that the number of each
representation in \eqref{eq:geomU1ACsol} must be a non-negative integer, which
gives us the conditions \eqref{eq:q2-conditions}.

Thus, it is not guaranteed from these considerations that every
anomaly-free $\U(1)$ model with matter of charge at most $\pm2$ can
be unHiggsed to a consistent $\SU(2)$ model. In particular, even
if we assume that $\tilde{n}$ is an integer, there are solutions of the second
and third conditions in \eqref{eq:q2-conditions} (such as $\tilde{n} = 2,
\gamma = 1$) that do not take the form of \eqref{eq:SU2-matching}. On the
other hand, if we assume that $\tilde{b}$ is an even element of the string
charge lattice $\Gamma$ (i.e., $\tilde{b} / 2 \in \Gamma$), then $\tilde{n}
\in 4 \Z$ and there is always an anomaly-free $\SU(2)$ model with $b =
\tilde{b} / 2$ that gives the corresponding $\U(1)$ model with charges $\pm1,
\pm2$ after Higgsing, up to the condition that the $\U(1)$ theory has
sufficient uncharged scalars to match the charge 0 contribution from
\eqref{eq:breaking-symmetric}. Recently the condition that $\tilde{b}$ is an
even element of the string charge lattice was proven under some mild
assumptions in \cite{MonnierMooreParkQuantization}. We discuss this condition
further in the context of F-theory constructions and the swampland in
sections~\ref{subsec:charge2-fTheory} and \ref{sec:swampland}.


\subsection{$T = 0$}\label{subsec:charge2-T0}

We now specialize to the case of models with no tensor multiplets, $T = 0$. In
this case, the vectors $a$, $b$, and $\tilde{b}$ simply become numbers. In
particular, we have $a = -3$; in order for the kinetic terms of the gauge
fields to have the appropriate sign, $b$ must be positive, and so the $a \cdot
b$ condition fixes the sign of $a$ to be negative. In this case, it is useful
to use the AC conditions in the form \eqref{eq:U1AC} rather than
\eqref{eq:geomU1AC}.

We restrict the maximum $\U(1)$ charge to be $q = \pm2$. The abelian AC
conditions \eqref{eq:U1AC} give us
\begin{subequations}
\label{eq:charge2-T0U1AC}
\begin{align}
18 \tilde{b} &= x_1 + 4 x_2\,, \label{eq:charge2-T0U1AC1} \\
3 \tilde{b}^2 &= x_1 + 16 x_2\,, \label{eq:charge2-T0U1AC2}
\end{align}
\end{subequations}
yielding models of the form
\begin{equation}
\label{eq:charge2-T0U1sol}
\tilde{b} \left(24 - \tilde{b}\right) \times (\bm{\pm1})
    + \frac{\tilde{b} \left(\tilde{b} - 6\right)}{4} \times (\bm{\pm2})\,.
\end{equation}
Note that requiring $\frac{\tilde{b} (\tilde{b} - 6)}{4} \in \N =\Z_{\ge 0}$
implies that $\tilde{b}$ is an even integer. We see that $\tilde{b} \ge 0$
from \refeqn~\eqref{eq:charge2-T0U1AC1}, and $6 \le \tilde{b} \le 24$ from the
condition that the numbers of fields of charge $\pm1$ and $\pm2$ is
non-negative. In this set of models, we have $H = x_0 + x_1 + x_2$,
$V = 1$, and $T = 0$, so the gravitational condition becomes $274 \ge x_1 +
x_2$, which all otherwise anomaly-consistent models satisfy. Note that in the
case $\tilde{b} = 24$, there are actually two distinct $\U(1)$ models with the
spectrum \eqref{eq:charge2-T0U1sol}, depending on the choice of unit charge;
we discuss this issue further below.

Now we consider $\SU(2)$ models containing only fundamentals and adjoints,
which we can Higgs to $\U(1)$ models with charges $q = \pm1, \pm2$ when there
is at least one adjoint field. Since $b$ is simply a number, we have $n =
b^2$. The corresponding value of $g$ is
\begin{equation}
g = \frac{(b - 1) (b - 2)}{2}\,.
\end{equation}
From \eqref{eq:geomSUACsol}, we find models of the form
\begin{equation}
\label{charge2-T0SU2sol}
2 b (12 - b) \times \ssyng{1} + \frac{(b - 1) (b - 2)}{2} \times
\ssyng{2}\,.
\end{equation}
Note that we must have $b \le 12$ in order to have a non-negative integer
number of each representation.

In general, we can include half-hypermultiplets of any quaternionic
(pseudoreal) irreducible representation. In this case, however, the
requirement that $b \in \Z$ (which follows from the integrality of $n$)
implies that $x_{\,\tyng{1}} \in \Z$. Furthermore, the gravitational condition
$273 = H - V + 29 T$ does not reduce the number of solutions in this case (as
in the $\U(1)$ case above): we have $H = x_0 + 2 x_{\,\tyng{1}} + 3
x_{\,\tyng{2}}$,
$V = 3$, and $T = 0$, so the condition becomes $276 \ge 2 x_{\,\tyng{1}} + 3
x_{\,\tyng{2}}$, which all otherwise anomaly-consistent models satisfy.

We can now Higgs the models \eqref{charge2-T0SU2sol} down to $\U(1)$ models
when $b \ge 3$ so that there is at least one adjoint field. We Higgs on a
charge that leaves the generator $T_3$ unbroken, as before, and find models of
the form
\begin{equation}
4 b (12 - b) \times (\bm{\pm1}) + b (b - 3) \times (\bm{\pm2})\,.
\end{equation}
Note that we have an exact matching between these models and those we found by
solving \eqref{eq:U1AC} if we take $\tilde{b} = 2 b$, once again matching the
expectation from \refsec~\ref{sec:relateAC}. This relies on the fact that
$\tilde{b}$ must be an even integer, as discussed previously.

Thus, we have determined that every anomaly-free $\U(1)$ massless spectrum
with $T = 0$ and matter of charge at most $\pm2$ can be unHiggsed to a model
with an $\su(2)$ gauge algebra. As we discuss in more detail in
\refsec~\ref{subsec:charge2-fTheory}, in fact each of the spectra of the form
\eqref{eq:charge2-T0U1sol} has an explicit realization in
F-theory. Except for some further subtleties that are relevant for the case
$\tilde{b} = 24, b = 12$, this indicates that in this simple context there is
no swampland.

For the case $\tilde{b} = 24$, there are some additional subtleties that are
useful to discuss. Note that, as discussed in
\refsec~\ref{subsec:background-addConstraints}, models with only
hypermultiplets of charge $q = \pm2$ are equivalent at the level of the
anomaly conditions to those with the same number of hypermultiplets of charge
$q = \pm1$. In the case under consideration here, the model with $\tilde{b} =
6$ has $x_1 = 108, x_2 = 0$ and the model with $\tilde{b} = 24$ has $x_1 = 0,
x_2 = 108$. However, each of these models can be unHiggsed in this case to an
anomaly-free $\SU(2)$ model with a distinct spectrum:
\begin{equation}
\begin{aligned}
54 \times \ssyng{1} + 1 \times \ssyng{2}
    &\longleftrightarrow 108 \times (\bm{\pm1})\,, \\
55 \times \ssyng{2}
    &\longleftrightarrow 108 \times (\bm{\pm2})\,.
\end{aligned}
\end{equation}
This illustrates the point mentioned earlier that models with massless fields
of charges that are all multiples of a non-unit integer $k$ can be physically
distinct from those that satisfy the condition that the GCD of the charges is
1. There is a further issue, however, which is that the $\tilde{b} = 24$ model
with 108 charges $q = \pm2$ can have two distinct realizations; in one, $q =
\pm1$ is the fundamental $\U(1)$ charge, and in the other, $q = \pm2$ is the
fundamental $\U(1)$ charge. The latter is still distinct from the model with
108 charges $\pm1$ and $\tilde{b} = 6$ due to the difference in values of
$\tilde{b}$. The two models with $\tilde{b} = 24$ unHiggs to two distinct
nonabelian models, the first of which has gauge group $\SU(2)$ and the second
of which has gauge group $\SO(3)$. As we discuss further in
\refsec~\ref{subsec:charge2-fTheory}, only the latter of these two models is
realized in F-theory (at least using known constructions).

\subsection{$T = 1$}\label{subsec:charge2-T1}

To classify $T = 1$ $\U(1)$ and $\SU(2)$ models from the supergravity point of
view, we must determine not only the possible values of $\tilde{n}, \gamma, n,
g$ that are consistent with anomaly cancellation but also which of these
values can be realized by vectors $a, b, \tilde{b}$ in the $(1, 1)$ signature
string charge lattice $\Gamma$. We know that the string charge lattice of a 6D
$\cN = 1$ theory must be unimodular~\cite{SeibergTaylorLattices}. There are
two unimodular lattices of signature $(1, 1)$, the odd lattice with inner
product
\begin{equation}
\Omega_1 = \mat[p]{1 & 0 \\ 0 & - 1}
\end{equation}
and the even lattice with inner product
\begin{equation}
\Omega_0 = U = \mat[p]{0 & 1 \\ 1 & 0}\,.
\end{equation}
We denote the corresponding string charge lattices by $\Gamma_1, \Gamma_0$.
For $\Gamma_1$, the only possible choice for $a$ so that $a \cdot a = 9 - T =
8$ is $-a = -a_1 = (3, -1)$, up to symmetries. For $\Gamma_0$ there are two
possibilities: $-a = -a_0 = (2, 2)$ and $-a = -a_0' = (4, 1)$. For brevity, we
denote the possible combinations of string charge lattice and $a$ vector by
$\Gamma_1$, $\Gamma_0$, and $\Gamma_0'$. Note that while both $\Gamma_1$ and
$\Gamma_0$ are realized in string theory with a variety of different choices
of positive cone (as we discuss further in the next section), there is no
known realization from string theory of models of type $\Gamma_0'$, i.e., the
lattice $\Gamma_0$ with $-a = (4, 1)$.

For each of the possible classes of models $\Gamma_1, \Gamma_0, \Gamma_0'$, we
have carried out a complete enumeration of the set of possible $\U(1)$ models,
including values of $\tilde{n}, \gamma$ and associated values of $\tilde{b}$.
For the 3 classes of models there are $195, 195, 314$ distinct allowed spectra
(i.e., integer non-negative values of $x_1, x_2$), respectively, that satisfy
the AC conditions, with $251, 383, 370$ possible distinct combinations of
spectra and $\tilde{b}$ values. Note that the number of spectra and
corresponding $\tilde{b}$ values includes a multiplicity for symmetries, i.e.,
$\tilde{b} = (b_1, b_2)$ and $\tilde{b} = (b_2, b_1)$ are both counted.

In this enumeration we have not imposed the positivity condition on
$\tilde{b}$; this is discussed further in the following section, but the basic
observation is that for the $\Gamma_0$ and $\Gamma_1$ lattices, there is a
choice of positive cone compatible with F-theory, generated by the elements
$(1, -1)$ and $(0, 1)$, which all anomaly-free solutions satisfy. (In fact,
all solutions on the even lattice are in the stricter positive cone generated
by $(1, 0)$ and $(0, 1)$.) Note that for $\Gamma_1$ and $\Gamma_0$ an even
$\tilde{b}$ always gives an even value of $x_2$, so an odd $x_2$ means that
$\tilde{b}$ is not even. This relation does not hold for $\Gamma_0',$ where we
can have $\tilde{b}$ even and $x_2$ odd.

The sets of possible $\U(1)$ spectra on the lattices $\Gamma_1$ and $\Gamma_0$
are exactly the same; each anomaly-free $\U(1)$ model on the even lattice has
a corresponding anomaly-free $\U(1)$ model (or several) on the odd lattice
with the same spectrum, and vice versa. There are, however, many $\U(1)$
models that satisfy the AC conditions that cannot be unHiggsed to an $\SU(2)$
model. In some cases, a spectrum can be associated with a value of $\tilde{b}$
on one lattice that admits an $\SU(2)$ unHiggsing, but cannot be unHiggsed on
the other lattice. For example, the spectrum
\begin{equation}
x_1 = 8\,, \quad x_2 = 142
\end{equation}
can be realized on the $\Gamma_1$ lattice by $\tilde{b} = (29, 9)$ or
$\tilde{b} = (43, -33)$, neither of which can be unHiggsed since $\tilde{b}$
is not even in either case. On the other hand, this spectrum can be realized
on the even lattice $\Gamma_0$ by $\tilde{b} = (10, 38)$, which can be
unHiggsed to the $\SU(2)$ model $x_{\,\tyng{1}} = 4, x_{\,\tyng{2}} = 72,
b = (5, 19)$.

Of the 251 (383) anomaly-free $\U(1)$ models (including spectrum and
$\tilde{b}$) on the lattice $\Gamma_1$ ($\Gamma_0$), 150 (280) models
(corresponding to 105 (140) spectra) are not unHiggsable to an $\SU(2)$ model.
Ninety of these spectra are not unHiggsable on either lattice. For many (40
spectra) of the models that cannot be unHiggsed on either lattice, the
unHiggsing cannot occur because they would violate the $\SU(2)$ gravitational
anomaly bound --- this condition is independent of $\tilde{b}$. For all the
remaining models that are not unHiggsable on either lattice (50 spectra),
$x_2$ is odd, implying that one component of $\tilde{b}$ is odd for both
choices of lattice. An example of such a model is
\begin{equation}
\label{eq:odd-example}
x_1 = 8\,, \quad x_2 = 97\,,
\end{equation}
which can be realized on the lattice $\Gamma_1$ by $\tilde{b} = (23, -3)$ and
on $\Gamma_0$ by $\tilde{b} = (13, 20)$. An example of a model that cannot be
unHiggsed due to the $\SU(2)$ gravitational anomaly bound is
\begin{equation}
\label{eq:gravity-example}
x_1 = 4\,, \quad x_2 = 236\,,
\end{equation}
which can be realized on the lattice $\Gamma_1$ by $\tilde{b} = (44, 26)$ and
on $\Gamma_0$ by $\tilde{b} = (9, 70)$. Note that there are 18 spectra that
cannot be unHiggsed on either lattice for both of these reasons.

There are 370 models on $\Gamma_0'$, giving 314 spectra, including all 195 of
the spectra from the other choices of lattice and $a$. They all satisfy
positivity for both cones described above. None of these models have F-theory
realizations, and we do not explore them further here.

\subsection{F-theory constraints and the swampland for charge $\pm1, \pm2$
models}\label{subsec:charge2-fTheory}

We now consider explicitly the set of $\U(1)$ models with $\abs{q} = 1, 2$
that can be realized in F-theory both directly as a $\U(1)$ Weierstrass model
and indirectly through Higgsing of an $\SU(2)$ model. This section assumes
some basic familiarity with F-theory; for more background see
\cite{MorrisonVafaI,MorrisonVafaII,KumarMorrisonTaylorGlobalAspects}.

In general, 6D supergravity models come from compactifying F-theory on a
complex surface $B$. The 6D string lattice is then associated with $H_{1,
1}(B, \Z)$, with the inner product arising from intersection of divisor
classes and the positive cone being the cone of effective divisors. For $T =
0$ the surface $B$ is $\bP^1$, and the canonical class is $-K = 3 H$. For $T =
1$ the surface $B$ is a Hirzebruch surface $\F_m, m \le 12$. When $m \ge 3$,
there are non-Higgsable nonabelian gauge groups everywhere in the moduli space
\cite{MorrisonTaylorClusters}, so a pure $\U(1)$ or $\SU(2)$ theory is only
possible for $m = 0, 1, 2$. In the cases $m = 0, 2$ we have the 6D charge
lattice $\Gamma_0$ with $-K = (2, 2)$ and for $m = 1$ we have $\Gamma_1$ with
$-K = (3, -1)$. In the case of $m = 1$, the effective cone on $\Gamma_1$ is
generated by the vectors $(0, 1)$ and $(1, -1)$, so for example $\tilde{b} =
(23, -3)$, as encountered in the example \eqref{eq:odd-example}, lies in the
positive cone of the corresponding 6D supergravity theory. For $m = 0, 2$, the
effective cone on $\Gamma_0$ is generated by $\{(1, 0), (0, 1)\}$ and $\{(1,
-1), (0, 1)\}$, respectively.

A 6D F-theory model with $\SU(2)$ gauge group realized on a divisor associated
with the vanishing locus of a function (really a section of a line bundle)
$\sigma = 0$ of divisor class $b = [\sigma]$ has a Weierstrass model
\begin{equation}
y^2 = x^3 + f x + g\,,
\end{equation}
where the discriminant vanishes to quadratic order in $\sigma$,
\begin{equation}
\label{eq:discriminant-sigma2}
\Delta = 4 f^3 + 27 g^2 = \sigma^2 (\tilde{\Delta})\,.
\end{equation}
A fairly general class of F-theory $\SU(2)$ models can be constructed by
choosing $f, g$ to have the form
\begin{subequations}
\label{eq:fg-2}
\begin{align}
f &= -\frac{1}{48} \phi^2 + f_1 \sigma + f_2 \sigma^2\,, \label{eq:f-2} \\
g &= \frac{1}{864} \phi^3 - \frac{1}{12} \phi f_1 \sigma
    + g_2\sigma^2\,, \label{eq:g-2}
\end{align}
\end{subequations}
which can be arranged either using the Tate tuning of a general Weierstrass
model \cite{MorrisonVafaII} or an order by order tuning
\cite{MorrisonTaylorMaS} so that the discriminant vanishes to order
$\sigma^2$. Here, $f, g$ are sections of line bundles $\cO(-4 K)$ and $\cO(-6
K)$. Thus, $f_1$ is a section of $\cO(-4 K - b)$, $f_2$ is a section of
$\cO(-4 K - 2 b)$, and so forth. If $-4 K - b$ is not effective, that is if $b
\le -4 K$ does not hold\footnote{Note that for divisors, the notation $A \le
B$ is used to indicate that $B - A$ is effective.}, then $f_1$, $f_2$, and
$g_2$ all vanish and $\Delta$ vanishes identically, which gives a globally
singular and unphysical F-theory model. So an $\SU(2)$ model of the form
\eqref{eq:fg-2} cannot exist unless $b \le -4 K$. Furthermore, the condition
on $b$ for the curve to have genus one or higher so that an adjoint Higgsing
is possible is that $-K \le b$. Thus, for a good F-theory model with an
$\SU(2)$ realized through the form \eqref{eq:fg-2} on a smooth divisor of at
least genus one, corresponding to the existence of at least one adjoint field,
we need
\begin{equation}
\label{eq:SU2-conditions}
-K \le b \le -4 K\,.
\end{equation}

The Morrison\nd{}Park direct construction gives, as in \eqref{eq:MP-b},
\begin{equation}
\tilde{b} = 2 (-K + [z])\,.
\end{equation}
In this construction, $[z]$ is effective, so $\tilde{b}$ is even, and $-2 K
\le \tilde{b}$. The construction also requires $\tilde{b} \le -8 K$, or as
in the nonabelian $\SU(2)$ model above the discriminant would vanish
identically, so we have
\begin{equation}
\label{eq:U1-conditions}
-2 K \le \tilde{b} \le -8 K\,.
\end{equation}
Furthermore, $b = \tilde{b} / 2 = -K + [z]$ can be explicitly described in the
Morrison\nd{}Park model as the locus that supports an $\SU(2)$ after an
explicit unHiggsing, and satisfies exactly the conditions
\eqref{eq:SU2-conditions}. In the absence of additional nonabelian gauge
groups, which might cause singularities upon unHiggsing of the $\U(1)$,
therefore, there is a precise necessary and sufficient condition for the
existence of a Morrison\nd{}Park $\U(1)$ model, which is equivalent to the
necessary and sufficient condition for the existence of an $\SU(2)$ model of
the form \eqref{eq:fg-2}. The condition is that there exist a divisor
$\tilde{b} = 2 b$ satisfying \eqref{eq:SU2-conditions} in the nonabelian case
and \eqref{eq:U1-conditions} in the abelian case.

Note that F-theory constructions of $\SU(2)$ models and $\U(1)$ models that do
not satisfy the conditions \eqref{eq:SU2-conditions} and
\eqref{eq:U1-conditions} are possible. In particular, as described in
\cite{KleversMorrisonRaghuramTaylorExotic}, if the divisor supporting the
$\SU(2)$ is singular, the leading terms in \eqref{eq:fg-2} can take a more
general ``non-UFD'' form where $\phi$ is not in the ring of global functions
on the divisor, so the upper bound from \eqref{eq:SU2-conditions} on $b$ does
not hold. Similar considerations motivated by the construction of $\U(1)$
models that can have higher charges \cite{Raghuram34} show that the upper
bound on $\tilde{b}$ from \eqref{eq:U1-conditions} also is not necessary when
the $\U(1)$ is realized in a fashion that does not take the Morrison\nd{}Park
form. Thus, these stronger constraints only hold for the standard F-theory
constructions described above and are not universal constraints on all
F-theory models. In fact, the ring of functions on $\sigma$ can fail globally
to be a UFD even when $\sigma$ is smooth, when the genus of $\sigma$ is
nonzero. For brevity in the subsequent discussion we refer to $\SU(2)$ models
of the form \eqref{eq:fg-2}, where the constraint \eqref{eq:SU2-conditions}
necessarily holds, as ``UFD $\SU(2)$ models,'' and models not of this form as
``non-UFD $\SU(2)$ models.''\footnote{We would like to thank Yinan Wang for
pointing out that some models of the form constructed in \cite{Raghuram34}
have no charges $\abs{q} > 2$. These models can be unHiggsed to non-UFD
$\SU(2)$ models, so that in fact there exist F-theory models that violate the
constraints \eqref{eq:SU2-conditions} and \eqref{eq:U1-conditions}. Such
exotic constructions exist even in some cases where the curve supporting the
gauge factor is smooth.} Note that for any $\SU(2)$ model in F-theory, the
constraint
\begin{equation}
\label{eq:Kodaira}
b \le -6 K
\end{equation}
must be satisfied since $-12 K - \sigma^2$ must be effective for the
discriminant to have the form \eqref{eq:discriminant-sigma2}. This is a simple
example of the ``Kodaira constraint,'' discussed more generally in
\cite{KumarMorrisonTaylorGlobalAspects}.

For $T = 0$, $b$ for an $\SU(2)$ model is simply the degree of the curve on
which the $\SU(2)$ is tuned. The genus of this curve is $g = (b - 1) (b - 2) /
2$. The $\SU(2)$ constraint \eqref{eq:SU2-conditions} states that $3 \le b
\le 12$, and the corresponding condition for the Morrison\nd{}Park $\U(1)$
model is \eqref{eq:U1-conditions}. Because $a, \tilde{b}$ are simply numbers,
\eqref{eq:U1-conditions} is precisely equivalent to the condition
\eqref{eq:q2-last}, which states
\begin{equation}
-2 a \cdot \tilde{b} \le \tilde{b} \cdot \tilde{b}
    \le -8 a \cdot \tilde{b}\,.
\end{equation}
Since the anomaly cancellation conditions already impose the constraint that
$\tilde{b}$ is even in the case $T = 0$, we see that for $T = 0$ there is a
match between anomaly-allowed $\U(1)$ models with charges $q = \pm1, \pm2$ and
the models that can be constructed from F-theory.

As mentioned in \refsec~\ref{subsec:charge2-T0}, the $\tilde{b} = 24$ model
potentially has two distinct realizations, one with fundamental charge $q =
\pm1$ and one with fundamental charge $q = \pm2$. These two models unHiggs to
models with nonabelian gauge groups $\SU(2)$ and $\SO(3)$, respectively. Only
the latter of the two models has a natural realization in F-theory through the
Morrison\nd{}Park construction.\footnote{We would like to thank Ling Lin for
discussions on this point.} After the unHiggsing of the Morrison\nd{}Park
model, the resulting Weierstrass model has an $\su(2)$ gauge algebra. The
group is, however, $\SO(3) = \SU(2) / \Z_2$ in this case due to the presence
of $\Z_2$ torsion; this can be checked by noting that the Weierstrass model
takes a Tate form with $a_2, a_3, a_6 = 0$, which as shown in
\cite{MayrhoferMorrisonTillWeigandTorsion} gives the $\SO(3)$ gauge group.
This is natural, from the absence of massless fields that transform
nontrivially under the $\Z_2$ center of $\SU(2)$ in this case. On the other
hand, there is no clear reason from the low-energy theory that the massless
spectrum of the $\U(1)$ theory cannot consist only of charges $q = \pm2$ even
if the fundamental charge is $q = \pm1$, or that there cannot be an $\SU(2)$
model with only adjoints in the massless spectrum. While we do not have an
F-theory realization of these models, and suspect they do not exist, we also
cannot completely rule out an exotic non-UFD type F-theory realization. So
these models are likely but not certain candidates for the swampland. We
discuss some further aspects of this in connection with the ``completeness
hypothesis'' in \refsec~\ref{subsec:swampland-completeness}.

For larger values of $T$, the anomaly cancellation conditions from 6D
supergravity are weaker than the conditions from F-theory. These two sets of
conditions do seem, however to have some interesting parallels. For $\U(1)$
models with only charges $q = \pm1, \pm2$ and no additional gauge groups, the
AC conditions \eqref{eq:q2-conditions} impose the constraints
\begin{equation}
\label{eq:AC-constraints-2}
\tilde{b} \cdot \tilde{b} + 2 a \cdot \tilde{b} \in 4 \Z\,, \quad
    -2 a \cdot \tilde{b} \le \tilde{b} \cdot \tilde{b}
    \le -8 a \cdot \tilde{b}\,.
\end{equation}
The existence of an F-theory model either from the Morrison\nd{}Park
construction or from Higgsing a UFD $\SU(2)$ model with only fundamentals and
adjoints imposes the constraints
\begin{equation}
\label{eq:F-constraints-2}
\tilde{b} \in 2 \Z, \quad -2 a \le \tilde{b} \le -8 a\,.
\end{equation}
Similarly, the anomaly constraints on an $\SU(2)$ model with only fundamental
and adjoint charges, no other gauge groups, and at least one adjoint field are
\begin{equation}
\label{eq:AC-constraints-SU2}
-a \cdot b \le b \cdot b \le -4 a \cdot b\,,
\end{equation}
where the first inequality comes from $g > 0$ and the second from the
constraint that the number of fundamental fields in \eqref{eq:SU2charge2} is
non-negative, $6 n + 16 (1 - g) \ge 0$. As in the abelian case, these
constraints are closely related to the conditions for the existence of a UFD
$\SU(2)$ model in F-theory,
\begin{equation}
\label{eq:F-constraints-SU2}
-a \le b \le -4 a\,.
\end{equation}
In both the abelian and nonabelian cases, the Morrison\nd{}Park and UFD
$\SU(2)$ F-theory constraints are strictly stronger than but closely parallel
to the constraints from anomalies. While the signature of the inner product is
indefinite, in the cases we consider here we must have $-a \cdot \tilde{b}
\ge 0$ and $\tilde{b} \cdot \tilde{b} \ge 0$. This follows directly from the
anomaly relations \eqref{eq:genericU1AC}. The analogous inequalities for $b$
also hold for $\SU(2)$ theories with at least one adjoint, from
\eqref{eq:genericSUAC}. Thus, while the anomaly cancellation conditions are a
necessary consequence of these F-theory conditions on specific models, not all
models satisfying the anomaly conditions satisfy these F-theory conditions.
The fact that exotic F-theory constructions can allow the constraints
\eqref{eq:F-constraints-2} and \eqref{eq:F-constraints-SU2} to be violated
makes it clear that these cannot really be low-energy constraints. But it does
suggest some interesting structure for certain generic classes of models, and
raises a question of whether Morrison\nd{}Park $\U(1)$ and UFD $\SU(2)$ models
have some special characteristic structure that can be identified in the
low-energy supergravity theory. We return to these questions in
\refsec~\ref{sec:swampland}.

The fact that the anomaly and F-theory constraints precisely agree for $q =
\pm1, \pm2$, $T = 0$ theories follows from the fact that the inner product is
simply a product of numbers in this case, with $\tilde{b}, b$ non-negative
integers and $\tilde{b}$ automatically even, so that the anomaly constraints
immediately imply the Morrison\nd{}Park $\U(1)$ and UFD $\SU(2)$ constraints.
For $T > 0$, the story becomes more complicated. From our analysis of $T = 1$
models, we have identified anomaly-free $\U(1)$ models for which $\tilde{b}$
is not even, and models for which $\tilde{b} \le -8 a$ is not satisfied, even
though in both cases the conditions \eqref{eq:AC-constraints-2} are satisfied.
If we assume, however, that $\tilde{b}$ must be even, it is unclear whether or
not the charge spectra with $T = 1$, $q = \pm1, \pm2$ that violate the
condition $\tilde{b} \le -8 a$ can be realized in F-theory. For most, but not
all, of those spectra that admit an unHiggsing to an anomaly-free $\SU(2)$
model, the Kodaira constraint \eqref{eq:Kodaira} is satisfied for some choice
of effective cone that is allowed from F-theory. Understanding which, if any,
of the models that violate the condition $\tilde{b} \le -8 a$ and can or
cannot be unHiggsed to an $\SU(2)$ model are in the swampland would require a
more complete understanding of non-UFD or non-Morrison\nd{}Park F-theory
constructions, likely building on the approach of \cite{Raghuram34}.

There are some spectra that arise, such as $(x_1, x_2) = (0, 150)$, which are
particularly interesting. This spectrum can be associated with $\tilde{b} =
(10, 40)$ on the even lattice $\Gamma_0$, which does not satisfy $\tilde{b}
\le -8 a$ on any effective cone from F-theory. Furthermore, the model can be
unHiggsed to an $\SU(2)$ model that satisfies the anomaly conditions, but has
$b = (5, 20)$ and violates the Kodaira bound for any effective cone from
F-theory. This model can also be realized on the odd lattice $\Gamma_1$, where
there is a similar issue. Thus, for both lattices, the resulting $\SU(2)$
model is in the ``swampland'' of theories that do not violate any known
quantum consistency condition based on the low-energy spectrum but definitely
cannot be realized in F-theory. In fact, there are eight other $\U(1)$ models,
such as $(x_1, x_2) = (4, 146)$ and $(x_1, x_2) = (12, 138)$, that have
$\SU(2)$ unHiggsings on one of $\Gamma_0$ or $\Gamma_1$ and that violate the
Kodaira bound for any effective cone from F-theory; each of the associated
$\SU(2)$ models is therefore in the swampland. It seems likely that these
$\U(1)$ models are also not realizable from F-theory and are in the swampland,
but we do not have any way of proving this at this time.

\section{Asymptotics and infinite anomaly-free charge families}
\label{sec:asymptotics}

In this section, we construct some infinite families of anomaly-free charge
spectra, and study the growth of the number of anomaly-free $\U(1)$ models in
the limit of large $\tilde{b}$, for $T = 0$. In this way, we can further
understand the proliferation of apparently consistent $\U(1)$ models at larger
charges.

\subsection{Large charge families}\label{subsec:asymptotics-infFam}


Before moving to asymptotics, we can directly address the question of whether
the total number of different spectra that satisfy the anomaly equations is
finite or infinite. As we saw at the end of
\refsec~\ref{subsec:background-addConstraints}, each solution to the abelian
AC equations trivially gives rise to an infinite family of models by scaling
the charges and the associated $\tilde{b}$, and thus we already know that the
number of distinct anomaly-free $\U(1)$ spectra is infinite. However, there
are additional nontrivial infinite families of $\U(1)$ models satisfying the
AC conditions. One such family is given by
\begin{equation}
\label{eq:infinite-3family}
54 \times (\bm{\pm q}) + 54 \times (\bm{\pm r})
    + 54 \times (\bm{\pm(q + r)})\,, \quad
    \tilde{b} = 6 \left(q^2 + q r + r^2\right)\,, \quad q, r \in \Z\,.
\end{equation}

\begin{figure}[!th]
\centering

\includegraphics[width=0.4\linewidth]{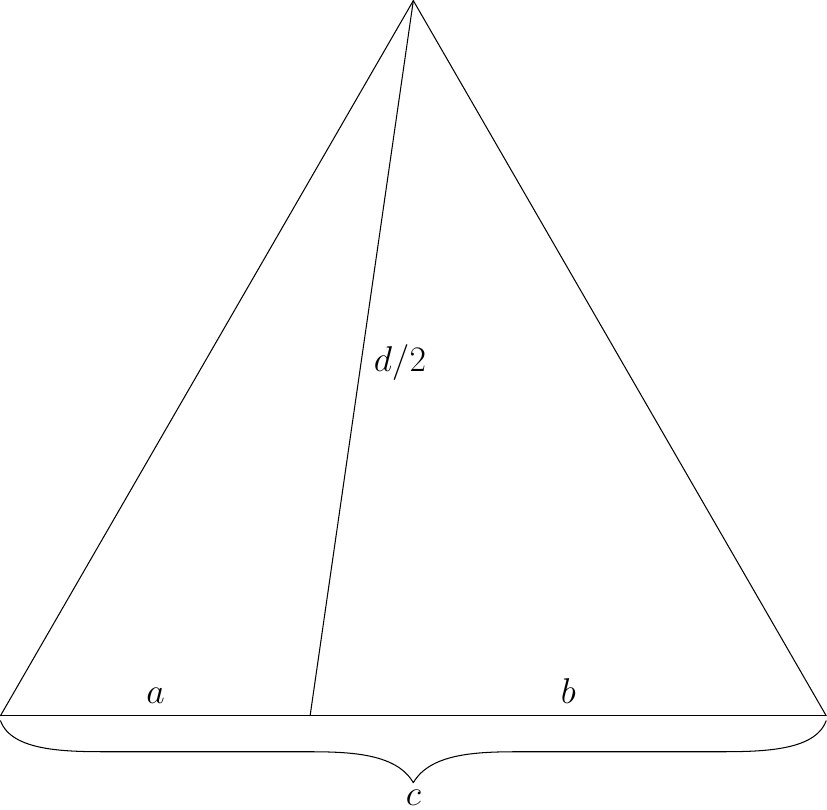}

\caption{An integer tuple $(a, b, c, d)$ such that there exists an equilateral
triangle of side length $c$ with an integral cevian of length $d / 2$ dividing
the side into integral parts, $a + b = c$. Such tuples produce anomaly-free
models of the form \eqref{eq:infinite-4family}.}
\label{fig:triangle}
\end{figure}

In analogy with \eqref{eq:infinite-3family}, we can similarly construct an
infinite family of four-charge models in the following fashion. Choose
distinct $m, n \in \Z_+$ with $n \le m / 2$, and define
\begin{equation}
\begin{aligned}
a &= m^2 - 2 m n\,, \\
b &= 2 m n - n^2\,, \\
c &= m^2 - n^2\,, \\
d &= 2 \left(m^2 - m n + n^2\right)\,.
\end{aligned}
\end{equation}
Then,
\begin{equation}
\label{eq:infinite-4family}
54 \times (\bm{\pm a}) + 54 \times (\bm{\pm b}) + 54 \times (\bm{\pm c})
    + 54 \times (\bm{\pm d})\,, \quad
\tilde{b} = 18 \left(m^2 - m n + n^2\right)^2
\end{equation}
is an anomaly-free model. Note that if $m \equiv -n \Mod 3$, then $\gcd(a, b,
c, d) = 3$, and $\gcd(a, b, c, d) = 1$ otherwise. These numbers have a
geometric interpretation, as shown in \reffig~\ref{fig:triangle}: $c$ is the
integral side length of an equilateral triangle such that there exists an
integral cevian (a line segment connecting a vertex of the triangle with any
point on the opposite side) of length $d / 2$ dividing $c$ into integral
parts, $a + b = c$. The set of $(a, b, c, d)$ generated in this way (after
dividing by the GCD, if necessary) is precisely the set of primitive integer
tuples satisfying this property.

These infinite families of anomaly-free $\U(1)$ spectra are particularly
interesting, because it is known that the number of distinct
anomaly-consistent spectra for nonabelian gauge groups and matter
representations in supergravity theories with $T < 9$ is finite. It is also
known that the number of consistent spectra (abelian or nonabelian) that can
arise from F-theory is finite. This indicates that there is an infinite
swampland of abelian models. The size of this swampland could be reduced
either by finding further quantum consistency constraints that rule out some
of these models, or by finding some other approach to string compactifications
that does not come from F-theory.

Note that there cannot be an infinite number of anomaly-free models with
charges below any given upper bound $q \le Q$, since the total number of
charged hypermultiplet fields is bounded by the gravitational anomaly
condition $\sum_q x_q \le 274$, and there are thus a finite number of spectra
with $q \le Q$.

Note also that issues related to the fundamental unit charge under the $\U(1)$
do not affect these infinite families. For example, the family
\eqref{eq:infinite-3family} with $q = 1$ and arbitrary $r$ gives an infinite
family with unit charge $1$.

\subsection{A continuous approximation}\label{subsec:asymptotics-contApprox}

The abelian AC conditions \eqref{eq:U1AC} can be written as
\begin{subequations}
\label{eq:asympRepsU1AC}
\begin{align}
\sum_{i = 1}^h q_i^2 = 18 \tilde{b}\,, \\
\sum_{i = 1}^h q_i^4 = 3 \tilde{b}^2\,,
\end{align}
\end{subequations}
where the sum runs over all charged matter representations, $i = 1, \dotsc,
h$, with the nonzero charges $q_i$ possibly degenerate.

To estimate the number of solutions of these equations for any given value of
$\tilde{b}$, we can consider this as a special case of the problem of finding
the number of solutions of the equations
\begin{subequations}
\label{eq:simplified-equations}
\begin{align}
\sum_{i = 1}^h q_i^2 &= B\,, \\
\sum_{i = 1}^h q_i^4 &= C\,,
\end{align}
\end{subequations}
where $B$ and $C$ are integers. We further define the quantity
\begin{equation}
\label{eq:x}
x = \frac{C}{B^2} = \frac{\sum_i q_i^4}{\left(\sum_i q_i^2\right)^2}\,.
\end{equation}
In our case of interest, $B = 18 \tilde{b}$, and $x = 1 / 108$, so $C = x B^2$
is automatically an integer for any integer $\tilde{b}$ and this value of $x$.

A simple scaling argument suggests that the number of solutions of the
equations~\eqref{eq:simplified-equations} should go roughly as $B^{(h - 6) /
2}$ for fixed $h$ at large $B$, for generic values of $x$ in the range $1 / h
\le x \le 1$. This can be seen as follows: the number of charge combinations
$q_i$ that satisfy $\sum_i q_i^2 \le B$ clearly scales as $B^{h / 2}$, as in
the continuum approximation it is the volume of an $h$-dimensional ball of
radius $\sqrt{B}$. The number of solutions that satisfy the equality $\sum_i
q_i^2 = B$ will go as the derivative of the number of solutions of the
inequality, so as $B^{(h - 2) / 2}$. The set of charge combinations $q_i$
gives a distribution on the set of $x$
that should approach a smooth distribution in the large $B$ limit, where the
limiting values $x = 1 / h, 1$ are realized by the charge combinations $(q, q,
\dotsc, q)$ and $(q, 0, 0, \dotsc, 0)$, respectively. The possible values of
$\left\lfloor x B^2\right\rceil$ (rounded integer value) are then distributed
across the range from $B^2 / h$ to $B^2$, so for any given $x$ the number of
solutions of \eqref{eq:simplified-equations} should scale as $B^{(h - 2) / 2}
/ B^2 = B^{(h - 6) / 2}$. This scaling can be understood geometrically as
arising from the intersection of the $(h - 1)$-dimensional sphere and quartic
hypersurface defined by the equations \eqref{eq:simplified-equations}.

This simple scaling argument suggests that when $h \ge 108$, the number of
solutions of \eqref{eq:asympRepsU1AC} will scale as $\tilde{b}^{(h - 6) / 2}$,
suggesting large infinite families of solutions, as the total number of
solutions up to an arbitrary $\tilde{b}$ should go as the integrated value,
$\tilde{b}^{(h - 4) / 2}$. By choosing simpler charge combinations, where a
large multiple of each of a smaller number of charges arises, this estimate
suggests, for example, that the number of solutions with $N$ copies each of
$m$ charges $q_1, \dotsc, q_m$ should scale as $\tilde{b}^{(m - 6) / 2}$, as
long as $108 \le m N \le 274$. The number of integrated solutions up to a
maximum value $\tilde{b}$ will then go as
\begin{equation}
\label{eq:asymptotic-estimate}
\cN_m^\text{approximate} \sim \tilde{b}^{(m - 4) / 2}\,.
\end{equation}
So, for example, there should be a logarithmically divergent number of
solutions with an equal number of four charges, a linearly divergent number of
solutions with an equal number of five charges, etc.

Note, however, that the infinite families identified in
\refsec~\ref{subsec:asymptotics-infFam} both exceed this estimate, with the
integrated number of models in the three-charge family scaling as
$\tilde{b}^1$, and the integrated number of models of the four-charge family
scaling as $\tilde{b}^{1 / 2}$. Indeed, this continuous scaling approximation
misses out on several features that modify this simplified analysis. First,
the continuous distribution on possible values of $x$ can have singular peaks
that enhance the number of solutions; second, number-theoretic considerations
can be relevant in affecting the number of solutions for particular values of
$B, x$.

We may also wish to restrict to cases where the $q_i$ do not all have a common
factor, in order to suppress the contribution of the trivial infinite families
produced by scaling of charges. This only affects the analysis by an overall
factor, and is only really relevant for small $h, m$.

The result of these features is that the scaling estimate just given is
generally an underestimate of the total number of solutions that can arise at
favorable combinations of charge multiplicities. To understand these issues
more clearly, we consider a simplified example in further detail.

\subsection{Two charges}\label{subsec:asymptotics-twoCharge}

As a simple toy example to illustrate the issues involved, we consider the
case of two charges, $h = 2$, and the system of equations
\begin{equation}
\label{eq:equations-2}
q^2 + r^2 = B\,, \quad q^4 + r^4 = x B^2\,.
\end{equation}
In the continuous approximation, we can estimate the number of solutions for
fixed $B, x$ by the integral
\begin{equation}
\label{eq:integral-2}
I_2(x) = \int_0^\infty \diff q \diff r \; \delta\left(q^2 + r^2 - B\right)
\delta\left(q^4 + r^4 - x B^2\right)\,.
\end{equation}
This gives the area in the $(q, r)$-plane that maps to a unit area in the
$\left(B, x B^2\right)$-plane, corresponding to the Jacobian of the
transformation between these spaces, which estimates the number of integer
solutions for $(q, r)$ corresponding to a given $(B, x)$ when $B \gg 1$. Using
the relation
\begin{equation}
\int \diff x \; g(x) \delta(f(x)) = \sum_{x_i} \frac{g(x_i)}{\abs{f'(x_i)}}\,,
\end{equation}
where the $x_i$ are the zeroes of $f(x)$ in the relevant integration range, we
can write \eqref{eq:integral-2} as
\begin{equation}
\begin{aligned}
I_2(x) &= \int_0^\infty \diff q \;
    \frac{\theta\left(B - q^2\right)}{2 \sqrt{B - q^2}}
    \delta\left(q^4 + \left(B - q^2\right)^2 - x B^2\right) \\
&= \frac{\theta(2 x - 1) \theta(1 - x)}{2 \sqrt{2} B^2
    \sqrt{2 x - 1} \sqrt{1 - x}}\,,
\end{aligned}
\end{equation}
with
\begin{equation}
\theta(x) = \begin{cases} 1\,, & x > 0 \\ 0\,, & x < 0\end{cases}
\end{equation}
the Heaviside step function. Here, we have assumed $B > 0$.

In accord with the general analysis of the preceding section, the number of
solutions in this continuous approximation thus scales as $B^{(2 - 6) / 2} =
B^{-2}$. Note, however, that the function $I_2(x)$ diverges at the values $x =
1 / 2, x = 1$. Thus, in the continuous approximation, while for generic values
of $x$ the number of solutions of the integer equations \eqref{eq:equations-2}
should scale as $B^{-2}$, the number of solutions for $x = 1 / 2$ appears to
diverge. From the fact that $C = x B^2$ is expected to be an integer, we can
estimate the effect of this divergence by integrating the divergent integrand
over the range from $C = B^2 / 2$ to $C = B^2 / 2 + 1$. Noting that $\diff C =
B^2 \diff x$, this yields
\begin{equation}
\int_{B^2 / 2}^{B^2 / 2 + 1} \frac{\diff C}{2 B^2 \sqrt{2 x - 1}}
    = \int_{1 / 2}^{1 / 2 + 1 / B^2} \frac{\diff x}{2 \sqrt{2 x - 1}}
    = \int_0^{1 / B^2} \frac{\diff\epsilon}{2 \sqrt{2 \epsilon}}
    = \frac{1}{\sqrt{2} B}\,,
\end{equation}
where we have approximated $\sqrt{2} \sqrt{1 - x} \approx 1$ over the entire
integration range. Thus, we may expect that the number of integer solutions of
\eqref{eq:equations-2} with $x = 1 / 2$ scales as $1 / B$, i.e., with an extra
factor of $B$. In the 3-charge case, a similar extra factor of $B$ contributes
to the scaling of the infinite family of charges \eqref{eq:infinite-3family},
which scales $\tilde{b}^{3 / 2}$ faster than the asymptotic estimate
\eqref{eq:asymptotic-estimate}.

\begin{figure}[!th]
\centering

\begin{minipage}{0.45\linewidth}
\includegraphics[width=\linewidth]{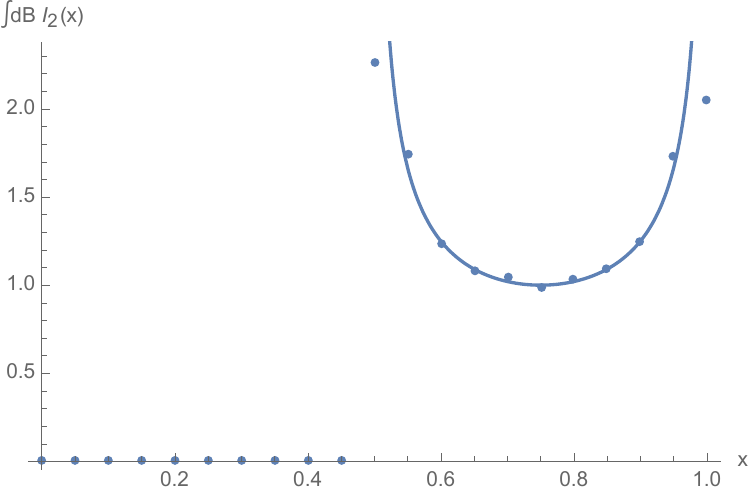}
\end{minipage}
\hspace*{0.05\linewidth}
\begin{minipage}{0.45\linewidth}
\includegraphics[width=\linewidth]{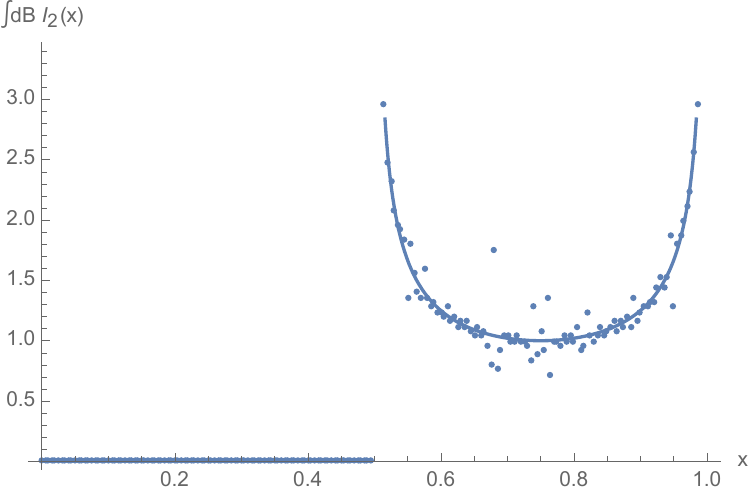}
\end{minipage}

\caption{The prediction for the distribution of values of $x = (q^4 + r^4) /
(q^2 + r^2)^2$ compared with numerical values for all $(q, r)$ with $q^2 + r^2
\le 10\,000$, where numerical values are averaged over bins of size $1 / 20$
(left) and $1 / 200$ (right). Irregularities in the right diagram indicate
number-theoretical patterns that become relevant for bins smaller than $1 /
\sqrt{B}$, such as the large peak at $x = 0.68$, as mentioned in the text.}
\label{fig:histograms}
\end{figure}

Let us consider the more precise number-theoretic aspects of this question
with $n = 2$. In fact, the solutions to \eqref{eq:equations-2} for $x = 1 / 2$
are those where $r = q, B = 2 q^2$, so the number of solutions with $q^2 + r^2
\le B$ and $x = 1 / 2$ is $\sqrt{B / 2}$, and the average number of solutions
of the equality goes as the derivative of this, or $1 / \sqrt{8 B}$. Thus, the
number of solutions at $x = 1 / 2$ exceeds even the above continuum estimate
adjusted to take account for the singularity. Note, however, that this family
of solutions seems to be unrelated to the singularity. There is another family
of solutions with $r = 2 q, B = 5 q^2, x = 17 / 25 = 0.68$ that provides
$\sim\sqrt{4 B / 5}$ solutions to the equations with the inequality, including
a factor of 2 to include solutions with $q = 2 r$. These families of solutions
arise for every relatively prime pair of integers $(a, b)$, where $(q, r) = (n
a, n b)$, and always give sets of order $\sqrt{B}$ of the total $B$ solutions
to the inequality $q^2 + r^2 \le B$. This pattern represents a breakdown of
the continuous approximation. Thus, we expect that a histogram of values of
$x$ for the set of all $q^2 + r^2 \le B$ will look fairly smooth if the bins
are larger than $1 / \sqrt{B}$, but exhibit number-theoretic irregularities
when the number of bins increases beyond $\sqrt{B}$. This is illustrated in
\reffig~\ref{fig:histograms}.

Related to this,
we may wish to restrict our attention to theories where $(q, r)$ are
relatively prime, to suppress the contribution of the trivial infinite
families seen in \refsec~\ref{subsec:background-addConstraints}. Note that in
the large $B$ limit, using the fact that the probability that a given pair of
integers is relatively prime is $6 / \pi^2$, the overall distribution should
be multiplied by this factor when considering only relatively prime charges.



We can carry out a similar analysis for the case of three distinct
charges, four distinct charges, and so on. As in the case of two
charges, there can be enhancements to the na\"ive growth
coming from singular peaks in the distribution on values of $x$ as
well as from number-theoretic effects.



\subsection{Fixed number of distinct charges}
\label{subsec:asymptotics-distCharge}

From the preceding analyses, we can now consider the more general problem of
estimating the number of solutions of the anomaly equations with $k$ distinct
charges $q_1, \dotsc, q_k$ with multiplicities $n_1, \dotsc, n_k$. The
equations that must be solved are then
\begin{subequations}
\label{eq:multiplicity-equations}
\begin{align}
\sum_{i = 1}^k n_i q_i^2 &= B\,, \\
\sum_{i = 1}^k n_i q_i^4 &= \frac{1}{108} B^2\,,
\end{align}
\end{subequations}
where $B = 18 \tilde{b}$ is an integer divisible by 18. Since the total number
of charges $\sum_i n_i $ is bounded above by 274, for each $k$ there are a
finite number of possible relevant combinations of the multiplicities $n_i$.
The asymptotic number of solutions at large $B$ should then scale as the
number of solutions for any combination of multiplicities. For each distinct
combination of $k$ and multiplicities $n_i$, we have a similar set of
equations to those studied above. This suggests that for generic choices of
$k, n_i$ the integrated number of solutions will scale as $\tilde{b}^{(k -
4) / 2}$, and that at specific values associated with singularities like those
described above, the number of solutions will scale as $\tilde{b}^{(k - 2) /
2}$. Thus, we would expect based on these analyses that the total number of
solutions with 3 distinct charges with any $18 \tilde{b} \le B$ should scale
as $B^{1 / 2}$, the number of solutions with 4 distinct charges should scale
as $B$, and the number of solutions with 5 distinct charges should scale as
$B^{3 / 2}$. In this analysis, we have not accounted for number-theoretic
effects like those encountered in \refsec~\ref{subsec:asymptotics-twoCharge},
which are needed to account for the scaling of families like
\eqref{eq:infinite-3family}.

Summarizing the results of the analysis, the continuum estimates from
\refsec~\ref{subsec:asymptotics-contApprox} show that infinite families of
charges are expected, with increasing rates of growth in $\tilde{b}$ as the
number of distinct charges increases, and singularities and number theoretic
effects explain why certain infinite families like those identified in
\refsec~\ref{subsec:asymptotics-infFam} arise with slightly larger powers in
the growth rates than those predicted by the continuous approximation.

\section{Classifying models with charges $\abs{q} > 2$}
\label{sec:largerCharge}

As we saw in \refsec~\ref{sec:asymptotics}, there are, in fact, an infinite
number of anomaly-free spectra for low-energy $\U(1)$ models. As the number of
consistent spectra that can arise from F-theory is finite, this means that
there is an infinite swampland of distinct massless spectra for $\U(1)$
models. As discussed in the previous section, however, it is difficult to show
that any specific $\U(1)$ model does not have an F-theory realization. In
order to better understand this infinite swampland, therefore, it is
illustrative to consider particular cases. To this end, in this section we
first consider models with maximum $\U(1)$ charge $\abs{q} \le 3$, i.e., $x_q
= 0$ for $\abs{q} \ge 4$, and $T = 0$. We analyze the types of models that
arise, and note mechanisms by which we can identify models that are clearly
\emph{not} in the swampland. We then carry out the same analysis for models
with charges up to $\pm4$ and $\pm5$.

One of our primary tools for identifying a model as not being in the swampland
is unHiggsing; if a $\U(1)$ model can be unHiggsed to a nonabelian model that
is realizable by an F-theory compactification, then we can realize the $\U(1)$
model in F-theory by Higgsing the associated nonabelian F-theory model. In
particular, the most direct unHiggsing of a $\U(1)$ model is to an $\SU(2)$
model, and so we will classify the $\U(1)$ models we find, in part, by their
ability to be unHiggsed to an anomaly-free $\SU(2)$ model. We will also see
cases where models that cannot be unHiggsed to an anomaly-free $\SU(2)$ model
may be unHiggsed to an anomaly-free nonabelian model with a larger gauge
group. If this nonabelian model can be realized in F-theory, this similarly
identifies the model as not being part of the swampland. We additionally
present arguments that may help identify cases where a model cannot be
unHiggsed to any nonabelian model.

There are also cases, however, where we can directly construct a Weierstrass
model for a $\U(1)$ theory \cite{Raghuram34}, despite it apparently not being
unHiggsable to any anomaly-consistent nonabelian model. Thus, while having an
unHiggsing to a nonabelian model with an F-theory realization is sufficient to
show that a $\U(1)$ model is not in the swampland, showing that there does not
exist such an unHiggsing is not a guarantee that the model cannot be found in
F-theory, and is therefore not a complete criterion for identifying swampland
models.

Ultimately, we have not managed to identify any specific $\U(1)$ models for
which we can definitively say, ``This anomaly-free model cannot be realized in
F-theory and therefore is in the swampland,'' though we know that as we
continue to classify models with arbitrarily high charges, infinitely many (in
fact, cofinitely many) models must be in the swampland. Nonetheless, the
analysis here gives us an idea of the kinds of models that may arise as we
consider ever larger charges, and may give some hint as to what the potential
obstructions are to finding F-theory compactifications that realize them.

Throughout this section we restrict our attention to models with $T = 0$.

\subsection{Charge $\abs{q} \le 3$ $\U(1)$ models}
\label{subsec:largerCharge-U1}

Anomaly-free $\U(1)$ models with $T = 0$ can be found via exchanges of the
form \eqref{eq:U1charge3exchange} from the $\abs{q} \le 2$ models
\eqref{eq:charge2-T0U1sol}. From this, we find that anomaly-free $\U(1)$
models with charges $q = \pm1, \pm2, \pm3$ are of the form
\begin{equation}
\label{eq:charge3-U1sol}
\left[\tilde{b} \left(24 - \tilde{b}\right) + 15 x_3\right] \times (\bm{\pm1})
    + \left[\frac{\tilde{b} \left(\tilde{b} - 6\right)}{4}
    - 6 x_3\right] \times (\bm{\pm2}) + x_3 \times (\bm{\pm3})\,.
\end{equation}
The requirement that the number of hypermultiplets transforming in each
irreducible representation must be a non-negative integer tells us that
$\tilde{b}$ is a non-negative, even integer. The number of such models for
which $x_1, x_2, x_3$ are all non-negative integers is 260. The gravitational
condition in this case becomes $274 \ge x_1 + x_2 + x_3$, which reduces the
number of anomaly-free models to 245. Note that these numbers include models
with $x_3 = 0$.

\subsection{$\SU(2)$ models with 3-symmetric matter}
\label{subsec:largerCharge-SU2}

Now we consider $\SU(2)$ models containing fundamentals, adjoints, and
triple-symmetrics, which we can Higgs to $\U(1)$ models with charges $q =
\pm1, \pm2, \pm3$. We allow for half-hypermultiplets of both $\ssyng{1}$ and
$\ssyng{3}$, although we will once again find that all models have full
hypermultiplets in the fundamental. Anomaly-free models can be found via
exchanges of the form \eqref{eq:SU2sym3exchange} from the generic matter
models \eqref{eq:geomSUACsol}. Thus, we have models of the form
\begin{equation}
\left[2 b (12 - b) + 14 x_{\,\tyng{3}}\right] \times \ssyng{1}
    + \left[\frac{(b - 1) (b - 2)}{2} - 6 x_{\,\tyng{3}}\right]
        \times \ssyng{2}
    + x_{\,\tyng{3}} \times \ssyng{3}\,.
\end{equation}
The number of such models for which $b, 2 x_{\,\tyng{1}}, x_{\,\tyng{2}}, 2
x_{\,\tyng{3}}$ are all non-negative integers and $x_{\,\tyng{2}} \ge 1$ is
223 (we require the number of adjoints to be at least 1 so that we can Higgs
on an adjoint). In this case, the gravitational condition becomes $276 \ge 2
x_{\,\tyng{1}} + 3 x_{\,\tyng{2}} + 4 x_{\,\tyng{3}}$, which reduces the
number of anomaly-free models to 199. Note that these numbers include models
with $x_{\,\tyng{3}} = 0$.

We can see already (since $245 > 199$) that there are $\U(1)$ models with
$\abs{q} \le 3$ that cannot be obtained by Higgsing an anomaly-free $\SU(2)$
model. To explore this further, we need to look at the Higgsing of these
$\SU(2)$ models down to $\U(1)$ models. We again Higgs on an adjoint charge
leaving the generator $T_3$ unbroken, which gives models of the form
\begin{equation}
\label{eq:T0higgsedSU2max3}
\left[4 b (12 - b) + 30 x_{\,\tyng{3}}\right] \times (\bm{\pm1})
    + \left[b (b - 3) - 12 x_{\,\tyng{3}}\right] \times (\bm{\pm2})
    + 2 x_{\,\tyng{3}} \times (\bm{\pm3})\,.
\end{equation}
As in \refsec~\ref{subsec:charge2-T0}, these solutions match the form coming
from the abelian AC conditions by taking $\tilde{b} = 2 b$, as well as $x_3 =
2 x_{\,\tyng{3}}$.

\subsection{Classification of charge $\abs{q} \le 3$ $\U(1)$ models}
\label{subsec:largerCharge-models}

We can now divide the anomaly-free $\U(1)$ models with charge $\abs{q} \le 3$
and integer non-negative spectra into 
five classes, grouped into 
two broad types, as shown in \reftab~\ref{tab:q3classes}.

The total set of $\U(1)$ models we are considering are those that satisfy the
AC conditions, including the gravitational bound. In classifying these models,
it is useful to distinguish the gravitational bound from the other AC
conditions, which we will loosely refer to as non-gravitational (NG), despite
the fact that they include the mixed gauge--gravitational anomaly condition.
Along with the NG conditions, we include the requirement that the number of
hypermultiplets transforming under any representation must be a non-negative
integer. We collectively refer to these conditions by NGIN (Non-Gravitational,
Integrality, Non-negativity). Note that we will also use the initialism NGIN
to refer to these conditions for $\SU(N)$, weakening the integrality condition
to allow quaternionic irreducible representations to have
half-hypermultiplets.

Each anomaly-free $\U(1)$ model can be na\"ively unHiggsed to some
(potentially inconsistent) $\SU(2)$ model; we say that a $\U(1)$ model
satisfies the $\SU(2)$ NGIN conditions if its corresponding unHiggsed $\SU(2)$
satisfies these conditions. Similarly, we say that a $\U(1)$ model satisfies
the $\SU(2)$ gravitational bound if its corresponding unHiggsed $\SU(2)$
satisfies this bound. Note that we consider all models with $\abs{q} \le 3$,
which includes all models with $\abs{q} \le 2$.

\begin{table}[!th]
\centering

\begin{tabular}{ccccc} \toprule
Type & \# & Exchangeable & $\SU(2)$ NGIN & $\SU(2)$ Grav 
\\ \midrule
\multirow{2}{*}{UnHiggsable} & 72 & \checkmark & \checkmark & \checkmark
\\
& 127 & & \checkmark & \checkmark 
\\ \midrule
\multirow{3}{*}{Non-unHiggsable} & 37 & & & \checkmark 
\\
& 2 & \checkmark & \checkmark & 
\\
& 7 & & \checkmark & 
\\ \bottomrule
%
\end{tabular}

\caption{Classes of $\abs{q} \le 3$ models satisfying the $\U(1)$ AC
conditions, as well as the integrality and non-negativity conditions. Classes
are grouped into two types: those unHiggsable to an anomaly-free $\SU(2)$
model and those that are not.
}
\label{tab:q3classes}
\end{table}

\subsubsection{UnHiggsable $\U(1)$s}\label{subsubsec:largerCharge-models-uH}

First, we consider the models that can be unHiggsed to an anomaly-free
$\SU(2)$ model. These are subdivided into two classes.

\paragraph{Exchangeable, unHiggsable $\U(1)$s}
First, there are those models that can be reached from a $\U(1)$ model with
$\abs{q} \le 2$ using exchanges of the form given in
\eqref{eq:U1charge3exchange}. This $\U(1)$ exchange corresponds exactly to the
$\SU(2)$ exchange \eqref{eq:SU2sym3exchange}, and in this case none of the
corresponding $\SU(2)$ exchanges results in a half-integer number of adjoints,
thus allowing all $\U(1)$ models in this class to be unHiggsed to $\SU(2)$.
There are 72 such models. Note that these models all satisfy the $\SU(2)$ Tate
bound $\tilde{b} \le 24$; in fact, the exchangeable $\U(1)$ models (which
include two non-unHiggsable models
, as discussed in 
the next section) are precisely those models that satisfy the $\SU(2)$ Tate
bound.

\paragraph{Non-exchangeable, unHiggsable $\U(1)$s}
Second, we have models that cannot be reached by $\U(1)$ exchanges from an
acceptable $\abs{q} \le 2$ model, but can nevertheless be unHiggsed to an
anomaly-free $\SU(2)$ model. Such models have $\tilde{b} (24 - \tilde{b}) < 0$
in \eqref{eq:charge3-U1sol}, but have a positive number of $(\bm{\pm1})$
representations due to the value of $x_3$; thus, they satisfy AC, but cannot
be exchanged to a maximum charge $q = \pm2$ model because the result would
have a negative number of $(\bm{\pm1})$ representations. There are 127 such
models.

\subsubsection{Non-unHiggsable $\U(1)$s}
\label{subsubsec:largerCharge-models-nuH}

Next, we consider the models that cannot be unHiggsed to an anomaly-free
$\SU(2)$ model. These are further subdivided into three classes. Such models
are the most interesting because we have the least understanding of how to
show whether or not they can be realized in F-theory.

\paragraph{NGIN non-unHiggsable $\U(1)$s}
First, we have models that cannot be unHiggsed to an $\SU(2)$ model because
the resultant $\SU(2)$ model would have a negative number of fundamentals
(thus violating NGIN). There are 37 such $\U(1)$ models, listed in
\reftab~\ref{tab:T0U1max3nonnegNUH}. These models are precisely those that
have $x_3 > x_1$; we can see from \eqref{eq:T0higgsedSU2max3} that any $\U(1)$
model obtained by Higgsing an $\SU(2)$ must have $x_3 \le x_1$. None of these
models are $\U(1)$ exchangeable, and they all satisfy the $\SU(2)$
gravitational bound.

\begin{table}[!th]
\centering

\begin{tabular}{>{$}c<{$}>{$}c<{$}>{$}c<{$}>{$}c<{$}} \toprule
\tilde{b} & x_1 & x_2 & x_3 \\ \midrule
30        & 0   & 108 & 12 \\
32        & 14  & 100 & 18 \\
34        & 5   & 100 & 23 \\
34        & 20  & 94  & 24 \\
36        & 3   & 96  & 29 \\
36        & 18  & 90  & 30 \\
38        & 8   & 88  & 36 \\
38        & 23  & 82  & 37 \\
40        & 5   & 82  & 43 \\
40        & 20  & 76  & 44 \\
40        & 35  & 70  & 45 \\
42        & 9   & 72  & 51 \\
42        & 24  & 66  & 52 \\
42        & 39  & 60  & 53 \\
44        & 5   & 64  & 59 \\
44        & 20  & 58  & 60 \\
44        & 35  & 52  & 61 \\
44        & 50  & 46  & 62 \\
46        & 8   & 52  & 68 \\
46        & 23  & 46  & 69 \\
46        & 38  & 40  & 70 \\
46        & 53  & 34  & 71 \\
46        & 68  & 28  & 72 \\
48        & 3   & 42  & 77 \\
48        & 18  & 36  & 78 \\
48        & 33  & 30  & 79 \\
48        & 48  & 24  & 80 \\
48        & 63  & 18  & 81 \\
48        & 78  & 12  & 82 \\
50        & 5   & 28  & 87 \\
50        & 20  & 22  & 88 \\
50        & 35  & 16  & 89 \\
50        & 50  & 10  & 90 \\
50        & 65  & 4   & 91 \\
52        & 14  & 10  & 98 \\
52        & 29  & 4   & 99 \\
54        & 0   & 0   & 108 \\\bottomrule
\end{tabular}

\caption{Anomaly-free $T = 0$ $\U(1)$ models with maximum charge $q = \pm3$
that cannot be unHiggsed to an anomaly-free $\SU(2)$ model because of the NGIN
conditions.}
\label{tab:T0U1max3nonnegNUH}
\end{table}


We can obtain at least some of these models by Higgsing from larger gauge
groups. For example, consider Higgsing $\SU(3)$ down to a single $\U(1)$. To
do so, we must Higgs so that exactly one of the generators is left unbroken.
Take the Cartan generators $T_1, T_2$ to be such that the fundamental has
charges $(1, 1), (-1, 0), (0, -1)$ and the adjoint has charges $(\pm2, \pm1),
(\pm1, \pm2), (\pm1, \mp1), 2 \times (0, 0)$ under these generators. We can
then Higgs on a charge such that $\SU(3)$ breaks down to $\U(1) \times \U(1)$
with $T_1, T_2$ the generators of the two factors, and then we can further
Higgs on the charge $(1, -1)$ so that only the generator $T_1 + T_2$ survives.
Under this Higgsing, we have
\begin{subequations}
\begin{align}
\ssyng{1} &\to 2 \times (\bm{\pm1}) + 1 \times (\bm{\pm2})\,, \\
\ssyng{2,1} &\to 4 \times (\bm{0}) + 4 \times (\bm{\pm3})\,.
\end{align}
\end{subequations}
We can then Higgs the anomaly-free $\SU(3)$ model
\begin{equation}
b = 8\colon \quad 24 \times \ssyng{1} + 21 \times \ssyng{2,1}
\end{equation}
to yield the $\U(1)$ model
\begin{equation}
\tilde{b} = 48\colon \quad 48 \times (\bm{\pm1}) + 24 \times (\bm{\pm2})
+ 80 \times (\bm{\pm3})\,,
\end{equation}
which appears in the list of 37 $\U(1)$ models that cannot be unHiggsed to
$\SU(2)$ models due to the NGIN conditions. Note that $\tilde{b} = 6 b$, as
expected from \refsec~\ref{sec:relateAC}.

In general, for some larger gauge group of rank $r$, we can Higgs down to a
single $\U(1)$ by choosing some $r - 1$ linearly independent roots (adjoint
charges) in the root space and projecting these out to see how the irreducible
representations split.

\paragraph{Exchangeable, gravity non-unHiggsable $\U(1)$s}
Second, we have exchangeable $\U(1)$ models that satisfy the $\SU(2)$ NGIN
conditions, but cannot be unHiggsed to an anomaly-free $\SU(2)$ model due to
the gravitational bound. In other words, these models do not contain enough
uncharged scalars to perform the unHiggsing. There are two such models, listed
in \reftab~\ref{tab:T0U1max3gravNUHex}.

\begin{table}[!th]
\centering

\begin{tabular}{>{$}c<{$}>{$}c<{$}>{$}c<{$}>{$}c<{$}} \toprule
\tilde{b} & x_1 & x_2 & x_3 \\ \midrule
22        & 254 & 4   & 14 \\
24        & 240 & 12  & 16 \\ \bottomrule
\end{tabular}

\caption{Anomaly-free, exchangeable $T = 0$ $\U(1)$ models with $\abs{q} \le
3$ that cannot be unHiggsed to an anomaly-free $\SU(2)$ model because of the
gravitational bound.}
\label{tab:T0U1max3gravNUHex}
\end{table}

The second of these models, with $\tilde{b} = 24$, can be found in F-theory by
a non-UFD construction, as described in \cite{Raghuram34}, with $[\eta_a]
\cdot [\eta_b] = 16$.

\paragraph{Non-exchangeable, gravity non-unHiggsable $\U(1)$s}
Finally, we have non-exchangeable $\U(1)$ models that nevertheless satisfy the
$\SU(2)$ NGIN conditions, but cannot be unHiggsed to an anomaly-free $\SU(2)$
model due to the gravitational bound. There are seven such models, listed in
\reftab~\ref{tab:T0U1max3gravNUH}.

\begin{table}[!th]
\centering

\begin{tabular}{>{$}c<{$}>{$}c<{$}>{$}c<{$}>{$}c<{$}} \toprule
\tilde{b} & x_1 & x_2 & x_3 \\ \midrule
26        & 233 & 16  & 19 \\
28        & 233 & 16  & 23 \\
30        & 225 & 18  & 27 \\
32        & 224 & 16  & 32 \\
34        & 215 & 16  & 37 \\
36        & 213 & 12  & 43 \\
38        & 218 & 4   & 50 \\ \bottomrule
\end{tabular}

\caption{Anomaly-free, non-exchangeable $T = 0$ $\U(1)$ models with $\abs{q}
\le 3$ that cannot be unHiggsed to an anomaly-free $\SU(2)$ model because of
the gravitational condition.}
\label{tab:T0U1max3gravNUH}
\end{table}

We believe that the gravitational bound for $\SU(2)$ models that Higgs to
$\U(1)$ models with maximum charge $q = \pm 3$, $276 \ge 2 x_{\,\tyng{1}} + 3
x_{\,\tyng{2}} + 34 x_{\,\tyng{3}}$, is less strict than the gravitational
bounds associated with other nonabelian groups, which would imply that the
nine $\U(1)$ models in this and the previous class cannot be unHiggsed to any
nonabelian gauge group. We do not have a completely rigorous and exhaustive
proof of this assertion, but we give a partial analysis here.

We can see that some of these models cannot be unHiggsed to any $\SU(N)$ model
via adjoint Higgsings by considering the number of uncharged scalars. If we
begin with an $\SU(N)$ gauge group, then the ``quickest'' route (in the sense
of producing the fewest uncharged scalars) to a single $\U(1)$ unbroken
symmetry group is to Higgs on an adjoint charge that breaks to $\U(1)^{N -
1}$, and then Higgs $N - 2$ more times on charges coming from the original
$\SU(N)$ adjoints to break to a single $\U(1)$. This is the same Higgsing
procedure we saw in \refsec~\ref{sec:relateAC}. In order to carry out this
process, the original model must have at least two adjoints. The adjoint to
which we give a VEV in the first step splits into the would-be Goldstone
bosons that are eaten to give the gauge bosons mass, and the remaining modes
are uncharged; we must have at least one other adjoint in order to have
adjoint charges on which to Higgs away all but one of the remaining $\U(1)$s.

In the first step, the Cartan generators of the $\SU(N)$ are those that become
the generators of the unbroken $\U(1)^{N - 1}$. Every adjoint (including the
one given the VEV) contributes $N - 1$ uncharged scalars, which are precisely
Cartan charges. In each of the subsequent Higgsings, because we give a VEV to
a charge coming originally from an $\SU(N)$ adjoint, the negative of this
charge also exists; one of these becomes the would-be Goldstone that is eaten,
and the other becomes an uncharged scalar. Both such charges from any
additional adjoints become uncharged scalars.

Thus, in the entire process, we gain $x_\text{Adj} (N - 1)$ uncharged scalars
from Higgsing the first adjoint, and then $(2 x_\text{Adj} - 3) (N - 2)$
additional uncharged scalars from the subsequent Higgsings. Because
$x_\text{Adj} \ge 2$, the number of uncharged scalars must then be at least
\begin{equation}
x_0 \ge 2 (N - 1) + (N - 2) = 3 N - 4\,.
\end{equation}
The models
\begin{equation}
\begin{aligned}
\tilde{b} &= 22: \quad 254 \times (\bm{\pm1}) + 4 \times (\bm{\pm2})
    + 14 \times (\bm{\pm3})\,, \\
\tilde{b} &= 28: \quad 233 \times (\bm{\pm1}) + 16 \times (\bm{\pm2})
    + 23 \times (\bm{\pm3})\,, \\
\tilde{b} &= 32: \quad 224 \times (\bm{\pm1}) + 16 \times (\bm{\pm2})
    + 32 \times (\bm{\pm3})\,, \\
\tilde{b} &= 38: \quad 218 \times (\bm{\pm1}) + 4 \times (\bm{\pm2})
    + 50 \times (\bm{\pm3})\,,
\end{aligned}
\end{equation}
all have $x_0 = 2$ and are not unHiggsable to $\SU(2)$ models, and thus cannot
be reached by adjoint Higgsings of any $\SU(N)$ model, by the argument above.

\subsection{Charges $\abs{q} \le 4, 5$}\label{subsec:largerCharge-45}

We can carry out the same classification for $\U(1)$ models with charges
$\abs{q} \le 4$ and $\abs{q} \le 5$. We classify these $\U(1)$ models by which
of three properties they satisfy: anomaly equivalence with a $\abs{q} \le 2$
model (``$\U(1)$ exchangeability''), the $\SU(2)$ NGIN conditions, and the
$\SU(2)$ gravitational bound, 
as explained in \refsec~\ref{subsec:largerCharge-models}.
There are potentially eight distinct classifications. As we saw in
\refsec~\ref{subsec:largerCharge-models}, for $\abs{q} \le 3$ only five of
these classes actually contain models.

For $\abs{q} \le 4$ and $\abs{q} \le 5$, there exist models in each of the
eight possible classes, as shown in \reftab~\ref{tab:q45classes}. Note that
the number of exchangeable models is relatively small. Note also that as the
maximum charge grows, the number of models that satisfy NGIN but violate the
$\U(1)$ gravitational bound (not listed in \reftab~\ref{tab:q45classes}) grows
rapidly. Indeed, the gravitational anomaly bound for $\U(1)$ enforces a cutoff
on the allowed number of hypermultiplets, and so we expect that as we include
higher and higher charges, a vast majority of models satisfying NGIN will not
satisfy these bounds. Nonetheless, as discussed in the previous section, there
are still large infinite families of models that satisfy NGIN and also the
gravitational anomaly condition.

\begin{table}[!th]
\centering

\begin{tabular}{cccccc} \toprule
Type & $\#_4$ & $\#_5$ & Exchangeable & $\SU(2)$ NGIN & $\SU(2)$ Grav
\\ \midrule
\multirow{2}{*}{UnHiggsable} & 88 & 89 & \checkmark & \checkmark & \checkmark
\\
& 1\,157 & 11\,320 & & \checkmark & \checkmark 
\\ \midrule
\multirow{6}{*}{Non-unHiggsable} & 42 & 43 & \checkmark & & \checkmark
\\
& 3\,204 & 81\,320 & & & \checkmark 
\\
& 7 & 11 & \checkmark & \checkmark & 
\\
& 358 & 7\,344 & & \checkmark & 
\\
& 7 & 8 & \checkmark & & 
\\
& 426 & 22\,182 & & & 
\\ \bottomrule
%
\end{tabular}

\caption{Classes of $\abs{q} \le 4$ and $\abs{q} \le 5$ models satisfying the
$\U(1)$ AC conditions, as well as the integrality and non-negativity
conditions. Classes are grouped into two types: those unHiggsable to an
anomaly-free $\SU(2)$ model and those that are not.
}
\label{tab:q45classes}
\end{table}

\section{F-theory and the swampland}\label{sec:swampland}

One of the principal questions of this paper is to ascertain the scope of the
set of models that appear consistent from the point of view of anomalies and
other known quantum consistency conditions, but that are not realized in
F-theory. In this section we summarize some of the main results on such
``swampland'' models.

The main conclusion is that we have found there is an infinite swampland of
apparently consistent massless charge spectra for 6D $\U(1)$ supergravity
models that cannot be realized in F-theory. However, there is no single
specific $\U(1)$ model that we can identify for which we can prove
definitively that there is no F-theory realization.

\subsection{Anomaly constraints on $\U(1)$ and $\SU(N)$ models}
\label{subsec:swampland-AC}

For $\U(1)$ models with charges $q = \pm1, \pm2$, the anomaly constraints
imply the condition
\begin{equation}
\label{eq:12-constraints-again}
-2 a \cdot \tilde{b} \le \tilde{b} \cdot \tilde{b} \le -8a \cdot \tilde{b}\,,
\end{equation}
where $x \le y$ means that $y - x$ is in the ``effective'' positivity cone of
the theory. When the charges are in the range $q \le Q$, following the same
logic as described just following \eqref{eq:q2-last}, the anomaly equations
imply that any $\U(1)$ model must satisfy the constraint
\begin{equation}
\tilde{b} \cdot \tilde{b} \le -2 Q^2 a \cdot \tilde{b}\,.
\end{equation}
For $Q = 3$, this gives the constraint
\begin{equation}
\tilde{b} \cdot \tilde{b} \le 18 a \cdot \tilde{b}\,.
\end{equation}

For $\SU(2)$ models with only fundamental and adjoint matter, and at least one
adjoint field, the anomaly constraints imply the condition
\begin{equation}
\label{eq:su2-constraints}
-a \cdot b \le b \cdot b \le -4 a \cdot b\,.
\end{equation}
For $\SU(3)$, the analogous constraint is
\begin{equation}
\label{eq:su3-constraints}
-a \cdot b \le b \cdot b \le -3 a \cdot b\,.
\end{equation}

These constraints match nicely with the observation that Higgsing an $\SU(N)$
model on adjoint fields in a natural way gives a $\U(1)$ model with $\tilde{b}
= N (N - 1) b$.

\subsection{Evenness condition on $\tilde{b}$}\label{subsec:swampland-even}

For $T = 0$, the anomaly conditions imply that $\tilde{b}$ is an even integer.
For $T > 0$, the condition that $\tilde{b}$ be an even element of the charge
lattice is necessary for an F-theory realization, but is not imposed by
anomalies. Recently, \cite{MonnierMooreParkQuantization} showed that the
condition that $\tilde{b}$ is even follows from some fairly mild assumptions
(specifically, they assume that the theory can be compactified on any spin
manifold, and that there is an allowed gauge configuration in every
topological class). Thus, it seems that the evenness of $\tilde{b}$ is likely
a necessary condition for the consistency of the quantum gravity theory, and
models with non-even $\tilde{b}$ are presumably inconsistent and therefore not
really of interest or in the swampland.

\subsection{$T = 0$, $q = \pm1, \pm2$}\label{subsec:swampland-q12}

In the limited class of $\U(1)$ models where there are no tensor multiplets
($T = 0$), and charges are restricted to $q = \pm1, \pm2$, there is a precise
match between the set of models that satisfy the anomaly conditions and those
that can be realized in F-theory, except for the model with $x_1 = 0, x_2 =
108$, where only the model with fundamental $\U(1)$ charge $q = \pm2$ has a
known realization in F-theory. We discuss this exceptional case further below
in \refsec~\ref{subsec:swampland-completeness}. Other than this case, there is
no swampland for $T = 0$, $q = \pm1, \pm2$.

\subsection{Bounds from F-theory}\label{subsec:swampland-fBounds}

For $T > 0$ models with tensor multiplets, and for models with massless
fields having $\U(1)$ charges $\abs{q} > 2$, while all known F-theory models
satisfy the anomaly constraints, there are in general a variety of low-energy
models that have no known F-theory realization, and infinite classes of
anomaly-consistent models of which only a finite number can be realized in
F-theory. So in general F-theory constraints are stronger than the anomaly
constraints.

The only completely clear explicit constraint that we are aware of, however,
on $a$ and $b$ that holds for all F-theory models and is stronger than the
anomaly constraints is the ``Kodaira bound.'' In
\cite{KumarMorrisonTaylorGlobalAspects}, it was pointed out that any F-theory
model with an $\SU(N)$ gauge symmetry associated with a divisor $b$ (giving
the corresponding anomaly coefficient) must satisfy the bound
\begin{equation}
\label{eq:old-constraint}
N b \le -12 a\,,
\end{equation}
associated with the Kodaira condition that the discriminant is in the class
$-12 K$. The constraint \eqref{eq:old-constraint} also implies a relation
between the $R^2$ and $F^2$ terms in the low-energy theory. For $\SU(2),
\SU(3)$, the Kodaira bound becomes, respectively, $b \le -6 a$ and $b \le -4
a$, which have a similar form to, but are slightly weaker than, the upper
bounds on $b$ from \eqref{eq:su2-constraints} and \eqref{eq:su3-constraints}.
As discussed in \refsec~\ref{subsec:charge2-fTheory}, when $T > 1$ there are
anomaly-free $\SU(2)$ models that violate this Kodaira constraint and are
therefore in the swampland. There are associated $\U(1)$ models that arise
from Higgsing these $\SU(2)$ models, but we do not have any proof that these
cannot be realized in F-theory, though it seems likely that they are also in
the swampland.

\subsection{Morrison\nd{}Park $\U(1)$ bounds and UFD $\SU(N)$ bounds}
\label{subsec:swampland-MPUFDBounds}

F-theory models with $\U(1)$ or $\SU(N)$ gauge groups that are constructed
using standard methodology obey an interesting set of constraints closely
parallel to \eqref{eq:12-constraints-again}, \eqref{eq:su2-constraints}, and
\eqref{eq:su3-constraints}. F-theory models of the Morrison\nd{}Park form
\cite{MorrisonParkU1} all satisfy the condition
\begin{equation}
\label{eq:12-inequality-again}
 -2 a \le \tilde{b} \le -8 a\,,
\end{equation}
This condition implies \eqref{eq:12-constraints-again} in the physically
allowed cases, but is stronger than that condition. In fact, the evenness of
$\tilde{b}$ and the condition \eqref{eq:12-inequality-again} seem to be
sufficient conditions for an F-theory model to exist. The condition
\eqref{eq:12-inequality-again} is not, however, necessary for an F-theory
model to exist; exotic constructions using the methodology of
\cite{Raghuram34} can violate this condition, even for models with only
charges $q = \pm1, \pm2$.

For any $\SU(2)$ model that is realized in F-theory using the standard
Tate/UFD construction \eqref{eq:fg-2}, the inequality
\begin{equation}
\label{eq:2-bound-again}
 b \le -4 a
\end{equation}
must hold. For $T = 0$ this is equivalent to the anomaly constraint, but for
$T > 0$ this is a stronger condition. Similarly, for any UFD $\SU(3)$ model we
have the inequality
\begin{equation}
 b \le -3 a\,,
\end{equation}
which is again closely analogous to but stronger than the $\SU(3)$ anomaly
constraint \eqref{eq:su3-constraints}. As in the abelian case, it is very
suggestive that standard F-theory constructions give constraints that are so
closely related to the anomaly constraints. Nonetheless, models that violate
these UFD constraints can be constructed using the non-UFD methodology of
\cite{KleversMorrisonRaghuramTaylorExotic}.

The connection between these Morrison\nd{}Park and UFD constraints and the
anomaly constraints for abelian and nonabelian theories is quite intriguing.
It raises the natural question of whether there is some signature in the
low-energy theory of models that come from a UFD type construction (including
Morrison\nd{}Park) that would distinguish these models from more exotic models
that involve non-UFD algebraic structure in the F-theory realization. We leave
this as a provocative question for further research.

\subsection{Models with $\U(1)$ charges $\abs{q} > 2$}
\label{subsec:swampland-largerCharge}

We have identified infinite families of anomaly-consistent charge spectra for
$\U(1)$ theories. In these families, the charges become arbitrarily large;
indeed, there are only a finite number of anomaly-consistent models with any
fixed upper bound on the charge.

On the other hand, our understanding of F-theory models with higher $\U(1)$
charges is quite limited. Some special cases of $\U(1)$ models with charge
$\abs{q} = 3$ matter were studied in
\cite{KleversMayorgaPenaOehlmannPiraguaReuterToric,KleversTaylor3Sym}, and a
more systematic construction was described in \cite{Raghuram34}. At this point
only a few models with $\abs{q} = 4$ have been explicitly constructed
\cite{Raghuram34}, and no models with larger $q$ have been explicitly
constructed. The first examples with $q = \pm3$ matter were found in
\cite{KleversMayorgaPenaOehlmannPiraguaReuterToric} by systematically studying
the set of possible toric hypersurface fiber types. One approach to trying to
identify new models with larger $\U(1)$ charges would be to investigate more
fully the set of toric and complete intersection fiber types using methodology
such as that developed in
\cite{BraunGrimmKeitelGeometric,BraunGrimmKeitelComplete}. Another approach
would be to develop further the non-UFD approach to describing $\U(1)$ models
with higher charges, either directly or through Higgsing of nonabelian models
with exotic matter as in
\cite{Raghuram34,KleversMorrisonRaghuramTaylorExotic}.

In principle, models with up to $\abs{q} = 6$ must be realizable in F-theory;
for example, one can get a model with $\abs{q} = 6$ matter by tuning an
$\SU(6)$ on a quartic on $\bP^2$, which has three adjoints, and then Higgsing
on two of the adjoints to get a $\U(1)$. (A similar construction with $\SU(5)$
gives matter with $\abs{q} = 5$.) It is unknown whether any F-theory model can
in principle be constructed with $\U(1)$ matter having charge $\abs{q} > 6$,
but there is clearly a finite upper bound on $q$ as the number of distinct
spectra for F-theory models is itself finite. It does not seem straightforward
to get a charge greater than $\abs{q} = 6$ in F-theory by breaking an
$\SU(N)$, and there are some hints \cite{CollinucciFazziValandroMorrisonComm},
based on the approach to charge $\abs{q} = 2$ constructions in recent work
\cite{CollinucciFazziValandroFlops}, that it may be impossible to construct
consistent F-theory models with singularities carrying $\U(1)$ matter with
$\abs{q} > 6$. Further progress on the F-theory side is clearly needed to
understand these kinds of constraints more clearly.

\subsection{Constraints on the charge lattice and positivity}
\label{subsec:swampland-latticeConstraints}

We have focused here on explicit questions about the spectrum and constraints
related to anomaly constraints. There is a further set of constraints imposed
by F-theory, discussed partly in \cite{KumarMorrisonTaylorGlobalAspects},
which are a bit more difficult to assess from the point of view of the
low-energy theory. In particular, F-theory allows only certain positive cones
for charged strings, associated with the cone of effective divisors in the
F-theory compactifications space. At present we have no understanding of how
the positivity cone of the low-energy theory may be constrained, outside the
framework of F-theory. Even assuming that the positivity cone is fixed to be
one that comes from F-theory, F-theory has the constraint that whenever a
curve of self-intersection $C \cdot C \le -3$ is in the positive cone, there
must be an associated non-Higgsable nonabelian gauge group
\cite{MorrisonTaylorClusters}. No analogue of this argument is known from the
point of view of the low-energy theory, though it is natural to imagine that a
careful treatment of the worldvolume theory of the strings arising in the
charge lattice may provide such constraints. Here we have mostly left these
questions to the side, but they certainly must be addressed for a complete
understanding or clearing of the 6D supergravity swampland.

\subsection{The completeness hypothesis}
\label{subsec:swampland-completeness}

Another constraint on supergravity models that plays a role in this discussion
is the ``completeness hypothesis,'' which states that for a gauge theory
coupled to gravity, all possible charges of the gauge group are realized by
some state in the Hilbert space \cite{BanksSeibergSymm}. This is related to
the idea that there are no global symmetries in a quantum theory coupled to
gravity. While there are various arguments for these statements based on black
holes and related physics, and the statements have been proven in the AdS/CFT
context \cite{HarlowOoguri}, there is no complete proof of these statements in
the 6D flat-space supergravity context relevant for this paper. Thus, this is
a hypothesis or folk theorem in this context. Some of the models we encounter
should be ruled out by this hypothesis. In particular, we have discussed the
model with $T = 0$ and 108 massless fields of charge $q = \pm2$ with
$\tilde{b} = 24$, where the fundamental $\U(1)$ unit charge can be chosen to
be $q = \pm1$. This model appears not to be realized in F-theory, and at least
naively violates the completeness hypothesis. There is no principle we are
aware of, however, that rules out the existence of a massless spectrum with
these features in a theory that still has massive fields of charge $q = \pm1$.
Indeed, there are ``non-Higgsable'' $\U(1)$ models that can be realized in
F-theory \cite{MartiniTaylorSemitoric,MorrisonParkTaylorNHAbelian,WangU1s}
that have no massless charged fields in the low-energy spectrum. Nonetheless,
these models should contain massive charged fields, for example associated
with massive adjoint fields in an unHiggsing to $\SU(2)$. This shows that the
massless spectrum does not always completely determine the allowed charge
spectrum of the theory. In any case, models such as the $\tilde{b} = 24$ model
with fundamental charge $q = \pm1$ do not appear to have F-theory realizations
and seem to be in the swampland; a further study of these models and others
that do not obviously satisfy the completeness hypothesis seems like an
interesting direction for further research. Note, however, that models such as
the model with $x_1 = 0, x_2 = 108, x_3 = 12$ listed in
\reftab~\ref{tab:T0U1max3nonnegNUH}, which have no fields of charge $q = \pm1$
but for which the GCD of massless fields is 1, are not in tension with the
completeness hypothesis. For such a model, combining massless states of
charges $+3$ and $-2$ will generally give a massive state of charge $+1$. So
there is no reason that these spectra should not be possible for consistent
theories of quantum gravity.

\section{Conclusions}\label{sec:conclusion}

We have explored the anomaly constraints on 6D $\cN = 1$ supergravity models
with a single $\U(1)$ gauge factor. We have found that there are infinite
families of distinct spectra that satisfy the anomaly constraints and all
other known quantum consistency conditions for these theories. Since F-theory
can only give a finite set of distinct spectra, this means that there is an
infinite swampland of apparently consistent low-energy models with no UV
realization at present.

We have observed some constraints on certain standard F-theory constructions
that are closely parallel to anomaly constraints for $\U(1)$ models with small
charges and $\SU(N)$ models with small $N$. These F-theory constraints are
violated by more exotic constructions, but the close relationship between
these formulae suggests that there may be some physical meaning to these
F-theory constraints. For $\SU(N)$ models, the Kodaira condition $N b \le
-12 a$ provides a universal constraint on models that can arise from F-theory,
which is violated by some anomaly-consistent low-energy spectra. We do not
have a clear understanding of any analogous constraint from F-theory for
models with only an abelian $\U(1)$ gauge group.

Further work is needed in several directions to explore these questions. It is
important to determine whether some new consistency constraint may be
identified that would limit the set of consistent $\U(1)$ models to a finite
space of theories. It would be interesting to look for explicit arguments for
the inequalities expected from F-theory associated with the Kodaira
constraint, or the stronger constraints on certain classes of models that
parallel the low-energy anomaly constraints. All these constraints might be
understood in the low-energy theory by identifying inconsistencies in theories
where the higher derivative $R^2$ terms have coefficients that are too small
relative to the gauge couplings.

Finally, more work is needed on the F-theory side. We do not at this time have
any clear explicit construction of F-theory $\U(1)$ models with charges
$\abs{q} > 4$, or any systematic construction of models with charges $\abs{q}
> 3$. We do not know what the upper bound on $\U(1)$ charges in F-theory
models may be, although as discussed in the previous section, it may be
$\abs{q} \leq 6$. It would also be interesting to further explore the set of
consistent $\U(1)$ charge spectra in the presence of an additional nonabelian
gauge factor, relating anomaly constraints to some of the F-theory structure
studied in, e.g., \cite{LawrieSchaferNamekiWongU1}. Some progress in this
direction has been made in \cite{GrimmKapferKleversArithmetic,CveticLinU1}. A
variety of models of this type have been studied using explicit toric
constructions \cite{BraunGrimmKeitelGeometric,BraunGrimmKeitelComplete}, which
as in the pure $\U(1)$ case may help provide useful data for such analyses.

All of these questions suggest that further exploration of the matter fields
in $\U(1)$ models in 6D supergravity and in F-theory may be a rich arena for
further research in the near future.

\acknowledgments
We would like to thank Andr\'{e}s Collinucci, Jonathan Heckman, Craig Lawrie,
Ling Lin, David Morrison, Paul Oehlmann, Nikhil Raghuram, Roberto Valandro,
and Yinan Wang for helpful discussions. The authors are supported by DOE grant
DE-SC00012567.

\bibliographystyle{JHEP}
\bibliography{research}

\providecommand{\href}[2]{#2}\begingroup\raggedright\begin{thebibliography}{10}

\bibitem{GreenSchwarzWest6DAnom}
M.~B. Green, J.~H. Schwarz and P.~C. West, \emph{{Anomaly Free Chiral Theories
  in Six-Dimensions}},
  \href{http://dx.doi.org/10.1016/0550-3213(85)90222-6}{\emph{Nucl. Phys.}
  {\bfseries B254} (1985) 327--348}.

\bibitem{SagnottiGS}
A.~Sagnotti, \emph{{A Note on the Green-Schwarz mechanism in open string
  theories}}, \href{http://dx.doi.org/10.1016/0370-2693(92)90682-T}{\emph{Phys.
  Lett.} {\bfseries B294} (1992) 196--203},
  [\href{https://arxiv.org/abs/hep-th/9210127}{{\ttfamily hep-th/9210127}}].

\bibitem{KumarTaylorBound}
V.~Kumar and W.~Taylor, \emph{{A bound on 6D {$\mathcal{N} = 1$}
  supergravities}},
  \href{http://dx.doi.org/10.1088/1126-6708/2009/12/050}{\emph{Journal of High
  Energy Physics} {\bfseries 2009} (2009) 1--23},
  [\href{https://arxiv.org/abs/0910.1586}{{\ttfamily 0910.1586}}].

\bibitem{SeibergTaylorLattices}
N.~Seiberg and W.~Taylor, \emph{{Charge Lattices and Consistency of 6D
  Supergravity}}, \href{http://dx.doi.org/10.1007/JHEP06(2011)001}{\emph{JHEP}
  {\bfseries 06} (2011) 001},
  [\href{https://arxiv.org/abs/1103.0019}{{\ttfamily 1103.0019}}].

\bibitem{KumarTaylorStringUni6D}
V.~Kumar and W.~Taylor, \emph{{String Universality in Six Dimensions}},
  \href{http://dx.doi.org/10.4310/ATMP.2011.v15.n2.a3}{\emph{Adv. Theor. Math.
  Phys.} {\bfseries 15} (2011) 325--353},
  [\href{https://arxiv.org/abs/0906.0987}{{\ttfamily 0906.0987}}].

\bibitem{VafaSwamp}
C.~Vafa, \emph{{The String landscape and the swampland}},
  \href{https://arxiv.org/abs/hep-th/0509212}{{\ttfamily hep-th/0509212}}.

\bibitem{OoguriVafaSwamp}
H.~Ooguri and C.~Vafa, \emph{{On the Geometry of the String Landscape and the
  Swampland}},
  \href{http://dx.doi.org/10.1016/j.nuclphysb.2006.10.033}{\emph{Nucl. Phys.}
  {\bfseries B766} (2007) 21--33},
  [\href{https://arxiv.org/abs/hep-th/0605264}{{\ttfamily hep-th/0605264}}].

\bibitem{KumarMorrisonTaylorGlobalAspects}
V.~Kumar, D.~R. Morrison and W.~Taylor, \emph{{Global aspects of the space of
  6D {$\mathcal{N} = 1$} supergravities}},
  \href{http://dx.doi.org/10.1007/JHEP11(2010)118}{\emph{JHEP} {\bfseries 11}
  (2010) 118}, [\href{https://arxiv.org/abs/1008.1062}{{\ttfamily 1008.1062}}].

\bibitem{MonnierMooreParkQuantization}
S.~Monnier, G.~W. Moore and D.~S. Park, \emph{{Quantization of anomaly
  coefficients in 6D $\mathcal{N} = (1,0)$ supergravity}},
  \href{http://dx.doi.org/10.1007/JHEP02(2018)020}{\emph{JHEP} {\bfseries 02}
  (2018) 020}, [\href{https://arxiv.org/abs/1711.04777}{{\ttfamily
  1711.04777}}].

\bibitem{Erler6DAnom}
J.~Erler, \emph{{Anomaly cancellation in six-dimensions}},
  \href{http://dx.doi.org/10.1063/1.530885}{\emph{J. Math. Phys.} {\bfseries
  35} (1994) 1819--1833},
  [\href{https://arxiv.org/abs/hep-th/9304104}{{\ttfamily hep-th/9304104}}].

\bibitem{KumarParkTaylorT=0}
V.~Kumar, D.~S. Park and W.~Taylor, \emph{{6D supergravity without tensor
  multiplets}}, \href{http://dx.doi.org/10.1007/JHEP04(2011)080}{\emph{Journal
  of High Energy Physics} {\bfseries 2011} (2011) 1--33}.

\bibitem{ParkTaylorAbelian}
D.~S. Park and W.~Taylor, \emph{{Constraints on 6D Supergravity Theories with
  Abelian Gauge Symmetry}},
  \href{http://dx.doi.org/10.1007/JHEP01(2012)141}{\emph{Journal of High Energy
  Physics} {\bfseries 2012} (2012) 1--50}.

\bibitem{SadovFTheoryGreenSchwarz}
V.~Sadov, \emph{{Generalized Green-Schwarz mechanism in F-theory}},
  \href{http://dx.doi.org/10.1016/0370-2693(96)01134-3}{\emph{Phys. Lett.}
  {\bfseries B388} (1996) 45--50},
  [\href{https://arxiv.org/abs/hep-th/9606008}{{\ttfamily hep-th/9606008}}].

\bibitem{MorrisonParkU1}
D.~R. Morrison and D.~S. Park, \emph{{F-Theory and the Mordell-Weil Group of
  Elliptically-Fibered Calabi-Yau Threefolds}},
  \href{http://dx.doi.org/10.1007/JHEP10(2012)128}{\emph{JHEP} {\bfseries 10}
  (2012) 128}, [\href{https://arxiv.org/abs/1208.2695}{{\ttfamily 1208.2695}}].

\bibitem{ParkAnomalies}
D.~S. Park, \emph{{Anomaly Equations and Intersection Theory}},
  \href{http://dx.doi.org/10.1007/JHEP01(2012)093}{\emph{JHEP} {\bfseries 01}
  (2012) 093}, [\href{https://arxiv.org/abs/1111.2351}{{\ttfamily 1111.2351}}].

\bibitem{WittenSU2Anom}
E.~Witten, \emph{{An SU(2) Anomaly}},
  \href{http://dx.doi.org/10.1016/0370-2693(82)90728-6}{\emph{Phys. Lett.}
  {\bfseries 117B} (1982) 324--328}.

\bibitem{BershadskyVafaAnom6D}
M.~Bershadsky and C.~Vafa, \emph{{Global anomalies and geometric engineering of
  critical theories in six-dimensions}},
  \href{https://arxiv.org/abs/hep-th/9703167}{{\ttfamily hep-th/9703167}}.

\bibitem{SuzukiTachikawaAnom6D}
R.~Suzuki and Y.~Tachikawa, \emph{{More anomaly-free models of six-dimensional
  gauged supergravity}}, \href{http://dx.doi.org/10.1063/1.2209767}{\emph{J.
  Math. Phys.} {\bfseries 47} (2006) 062302},
  [\href{https://arxiv.org/abs/hep-th/0512019}{{\ttfamily hep-th/0512019}}].

\bibitem{KumarMorrisonTaylor6DSUGRA}
V.~Kumar, D.~R. Morrison and W.~Taylor, \emph{{Mapping 6D {$\mathcal{N} = 1$}
  supergravities to F-theory}},
  \href{http://dx.doi.org/10.1007/JHEP02(2010)099}{\emph{JHEP} {\bfseries 02}
  (2010) 099}, [\href{https://arxiv.org/abs/0911.3393}{{\ttfamily 0911.3393}}].

\bibitem{MorrisonTaylorMaS}
D.~R. Morrison and W.~Taylor, \emph{{Matter and singularities}},
  \href{http://dx.doi.org/10.1007/JHEP01(2012)022}{\emph{JHEP} {\bfseries 01}
  (2012) 022}, [\href{https://arxiv.org/abs/1106.3563}{{\ttfamily 1106.3563}}].

\bibitem{AndersonGrayRaghuramTaylorMiT}
L.~B. Anderson, J.~Gray, N.~Raghuram and W.~Taylor, \emph{{Matter in
  transition}}, \href{http://dx.doi.org/10.1007/JHEP04(2016)080}{\emph{JHEP}
  {\bfseries 04} (2016) 080},
  [\href{https://arxiv.org/abs/1512.05791}{{\ttfamily 1512.05791}}].

\bibitem{KleversMorrisonRaghuramTaylorExotic}
D.~Klevers, D.~R. Morrison, N.~Raghuram and W.~Taylor, \emph{{Exotic matter on
  singular divisors in F-theory}},
  \href{http://dx.doi.org/10.1007/JHEP11(2017)124}{\emph{JHEP} {\bfseries 11}
  (2017) 124}, [\href{https://arxiv.org/abs/1706.08194}{{\ttfamily
  1706.08194}}].

\bibitem{GrassiMorrisonAnomalies}
A.~Grassi and D.~R. Morrison, \emph{{Anomalies and the Euler characteristic of
  elliptic Calabi-Yau threefolds}},
  \href{http://dx.doi.org/10.4310/CNTP.2012.v6.n1.a2}{\emph{Commun. Num. Theor.
  Phys.} {\bfseries 6} (2012) 51--127},
  [\href{https://arxiv.org/abs/1109.0042}{{\ttfamily 1109.0042}}].

\bibitem{Raghuram34}
N.~Raghuram, \emph{{Abelian F-theory Models with Charge-3 and Charge-4
  Matter}},  \href{https://arxiv.org/abs/1711.03210}{{\ttfamily 1711.03210}}.

\bibitem{MartiniTaylorSemitoric}
G.~Martini and W.~Taylor, \emph{{6D F-theory models and elliptically fibered
  Calabi-Yau threefolds over semi-toric base surfaces}},
  \href{http://dx.doi.org/10.1007/JHEP06(2015)061}{\emph{JHEP} {\bfseries 06}
  (2015) 061}, [\href{https://arxiv.org/abs/1404.6300}{{\ttfamily 1404.6300}}].

\bibitem{MorrisonParkTaylorNHAbelian}
D.~R. Morrison, D.~S. Park and W.~Taylor, \emph{{Non-Higgsable abelian gauge
  symmetry and F-theory on fiber products of rational elliptic surfaces}},
  \href{https://arxiv.org/abs/1610.06929}{{\ttfamily 1610.06929}}.

\bibitem{MorrisonTaylorSections}
D.~R. Morrison and W.~Taylor, \emph{{Sections, multisections, and U(1) fields
  in F-theory}},  \href{https://arxiv.org/abs/1404.1527}{{\ttfamily
  1404.1527}}.

\bibitem{CveticKleversPiraguaTaylorU1U1}
M.~Cveti{\v c}, D.~Klevers, H.~Piragua and W.~Taylor, \emph{{General
  U(1){$\times$}U(1) F-theory compactifications and beyond: geometry of
  unHiggsings and novel matter structure}},
  \href{http://dx.doi.org/10.1007/JHEP11(2015)204}{\emph{JHEP} {\bfseries 11}
  (2015) 204}, [\href{https://arxiv.org/abs/1507.05954}{{\ttfamily
  1507.05954}}].

\bibitem{MorrisonVafaI}
D.~R. Morrison and C.~Vafa, \emph{{Compactifications of F theory on Calabi-Yau
  threefolds --- I}},
  \href{http://dx.doi.org/10.1016/0550-3213(96)00242-8}{\emph{Nucl. Phys.}
  {\bfseries B473} (1996) 74--92},
  [\href{https://arxiv.org/abs/hep-th/9602114}{{\ttfamily hep-th/9602114}}].

\bibitem{MorrisonVafaII}
D.~R. Morrison and C.~Vafa, \emph{{Compactifications of F theory on Calabi-Yau
  threefolds --- II}},
  \href{http://dx.doi.org/10.1016/0550-3213(96)00369-0}{\emph{Nucl. Phys.}
  {\bfseries B476} (1996) 437--469},
  [\href{https://arxiv.org/abs/hep-th/9603161}{{\ttfamily hep-th/9603161}}].

\bibitem{MorrisonTaylorClusters}
D.~R. Morrison and W.~Taylor, \emph{{Classifying bases for 6D F-theory
  models}}, \href{http://dx.doi.org/10.2478/s11534-012-0065-4}{\emph{Central
  Eur. J. Phys.} {\bfseries 10} (2012) 1072--1088},
  [\href{https://arxiv.org/abs/1201.1943}{{\ttfamily 1201.1943}}].

\bibitem{MayrhoferMorrisonTillWeigandTorsion}
C.~Mayrhofer, D.~R. Morrison, O.~Till and T.~Weigand, \emph{{Mordell-Weil
  Torsion and the Global Structure of Gauge Groups in F-theory}},
  \href{http://dx.doi.org/10.1007/JHEP10(2014)016}{\emph{JHEP} {\bfseries 10}
  (2014) 16}, [\href{https://arxiv.org/abs/1405.3656}{{\ttfamily 1405.3656}}].

\bibitem{KleversMayorgaPenaOehlmannPiraguaReuterToric}
D.~Klevers, D.~K. Mayorga~Pena, P.-K. Oehlmann, H.~Piragua and J.~Reuter,
  \emph{{F-Theory on all Toric Hypersurface Fibrations and its Higgs
  Branches}}, \href{http://dx.doi.org/10.1007/JHEP01(2015)142}{\emph{JHEP}
  {\bfseries 01} (2015) 142},
  [\href{https://arxiv.org/abs/1408.4808}{{\ttfamily 1408.4808}}].

\bibitem{KleversTaylor3Sym}
D.~Klevers and W.~Taylor, \emph{{Three-Index Symmetric Matter Representations
  of SU(2) in F-Theory from Non-Tate Form Weierstrass Models}},
  \href{http://dx.doi.org/10.1007/JHEP06(2016)171}{\emph{JHEP} {\bfseries 06}
  (2016) 171}, [\href{https://arxiv.org/abs/1604.01030}{{\ttfamily
  1604.01030}}].

\bibitem{BraunGrimmKeitelGeometric}
V.~Braun, T.~W. Grimm and J.~Keitel, \emph{{Geometric Engineering in Toric
  F-Theory and GUTs with U(1) Gauge Factors}},
  \href{http://dx.doi.org/10.1007/JHEP12(2013)069}{\emph{JHEP} {\bfseries 12}
  (2013) 069}, [\href{https://arxiv.org/abs/1306.0577}{{\ttfamily 1306.0577}}].

\bibitem{BraunGrimmKeitelComplete}
V.~Braun, T.~W. Grimm and J.~Keitel, \emph{{Complete Intersection Fibers in
  F-Theory}}, \href{http://dx.doi.org/10.1007/JHEP03(2015)125}{\emph{JHEP}
  {\bfseries 03} (2015) 125},
  [\href{https://arxiv.org/abs/1411.2615}{{\ttfamily 1411.2615}}].

\bibitem{CollinucciFazziValandroMorrisonComm}
A.~Collinucci, M.~Fazzi, R.~Valandro and D.~Morrison, \emph{Private
  communication},  2018.

\bibitem{CollinucciFazziValandroFlops}
A.~Collinucci, M.~Fazzi and R.~Valandro, \emph{{Geometric engineering on flops
  of length two}},  \href{https://arxiv.org/abs/1802.00813}{{\ttfamily
  1802.00813}}.

\bibitem{BanksSeibergSymm}
T.~Banks and N.~Seiberg, \emph{{Symmetries and Strings in Field Theory and
  Gravity}}, \href{http://dx.doi.org/10.1103/PhysRevD.83.084019}{\emph{Phys.
  Rev.} {\bfseries D83} (2011) 084019},
  [\href{https://arxiv.org/abs/1011.5120}{{\ttfamily 1011.5120}}].

\bibitem{HarlowOoguri}
D.~Harlow and H.~Ooguri, \emph{Symmetries in quantum field theory and quantum
  gravity},  to appear.

\bibitem{WangU1s}
Y.-N. Wang, \emph{{Tuned and Non-Higgsable U(1)s in F-theory}},
  \href{http://dx.doi.org/10.1007/JHEP03(2017)140}{\emph{JHEP} {\bfseries 03}
  (2017) 140}, [\href{https://arxiv.org/abs/1611.08665}{{\ttfamily
  1611.08665}}].

\bibitem{LawrieSchaferNamekiWongU1}
C.~Lawrie, S.~Schafer-Nameki and J.-M. Wong, \emph{{F-theory and All Things
  Rational: Surveying U(1) Symmetries with Rational Sections}},
  \href{http://dx.doi.org/10.1007/JHEP09(2015)144}{\emph{JHEP} {\bfseries 09}
  (2015) 144}, [\href{https://arxiv.org/abs/1504.05593}{{\ttfamily
  1504.05593}}].

\bibitem{GrimmKapferKleversArithmetic}
T.~W. Grimm, A.~Kapfer and D.~Klevers, \emph{{The Arithmetic of Elliptic
  Fibrations in Gauge Theories on a Circle}},
  \href{http://dx.doi.org/10.1007/JHEP06(2016)112}{\emph{JHEP} {\bfseries 06}
  (2016) 112}, [\href{https://arxiv.org/abs/1510.04281}{{\ttfamily
  1510.04281}}].

\bibitem{CveticLinU1}
M.~Cvetic and L.~Lin, \emph{{The Global Gauge Group Structure of F-theory
  Compactification with U(1)s}},
  \href{http://dx.doi.org/10.1007/JHEP01(2018)157}{\emph{JHEP} {\bfseries 01}
  (2018) 157}, [\href{https://arxiv.org/abs/1706.08521}{{\ttfamily
  1706.08521}}].

\end{thebibliography}\endgroup

\end{document}